\documentclass[sigconf]{acmart}

\usepackage{subcaption}
\usepackage{caption}
\usepackage{rotating} 
\usepackage{tikz}
\usepackage{tikzscale}
\usetikzlibrary{calc,arrows.meta}
\usepackage{makecell}
\usepackage{booktabs}
\usepackage{placeins}
\newcolumntype{L}{>{\raggedright\arraybackslash}X}
\newcolumntype{C}{>{\centering\arraybackslash}X}

\newcommand{\up}{\textcolor{orange}{$\uparrow$}}
\newcommand{\down}{\textcolor{blue}{$\downarrow$}}

\AtBeginDocument{%
  }

\setcopyright{acmlicensed}
\copyrightyear{2026}
\acmYear{2026}
\setcopyright{cc}
\setcctype{by}
\acmConference[CHI '26]{Proceedings of the 2026 CHI Conference on Human Factors in Computing Systems}{April 13--17, 2026}{Barcelona, Spain}
\acmBooktitle{Proceedings of the 2026 CHI Conference on Human Factors in Computing Systems (CHI '26), April 13--17, 2026, Barcelona, Spain}
\acmDOI{10.1145/3772318.3790976}
\acmISBN{979-8-4007-2278-3/2026/04}

\begin{document}

\title{Searching Through Complex Worlds: Visual Search and Spatial Regularity Memory in Mixed Reality}

\author{Lefan Lai}
\affiliation{
    \institution{School of Computer Science \\ The University of Sydney}
    \city{Sydney}
    \state{New South Wales}
    \country{Australia}
}
\email{lefan.lai.australia@gmail.com}

\author{Tinghui Li}
\affiliation{
    \institution{School of Computer Science \\ The University of Sydney}
    \city{Sydney}
    \state{New South Wales}
    \country{Australia}
}
\email{tinghui.li@sydney.edu.au}

\author{Zhanna Sarsenbayeva}
\affiliation{
    \institution{School of Computer Science \\ The University of Sydney}
    \city{Sydney}
    \state{New South Wales}
    \country{Australia}
}
\email{zhanna.sarsenbayeva@sydney.edu.au}

\author{Brandon Victor Syiem}
\affiliation{
    \institution{School of Computer Science \\ The University of Sydney}
    \city{Sydney}
    \state{New South Wales}
    \country{Australia}
}
\email{brandon.syiem@sydney.edu.au}

\begin{teaserfigure}
    \centerline{\includegraphics[width=\textwidth]{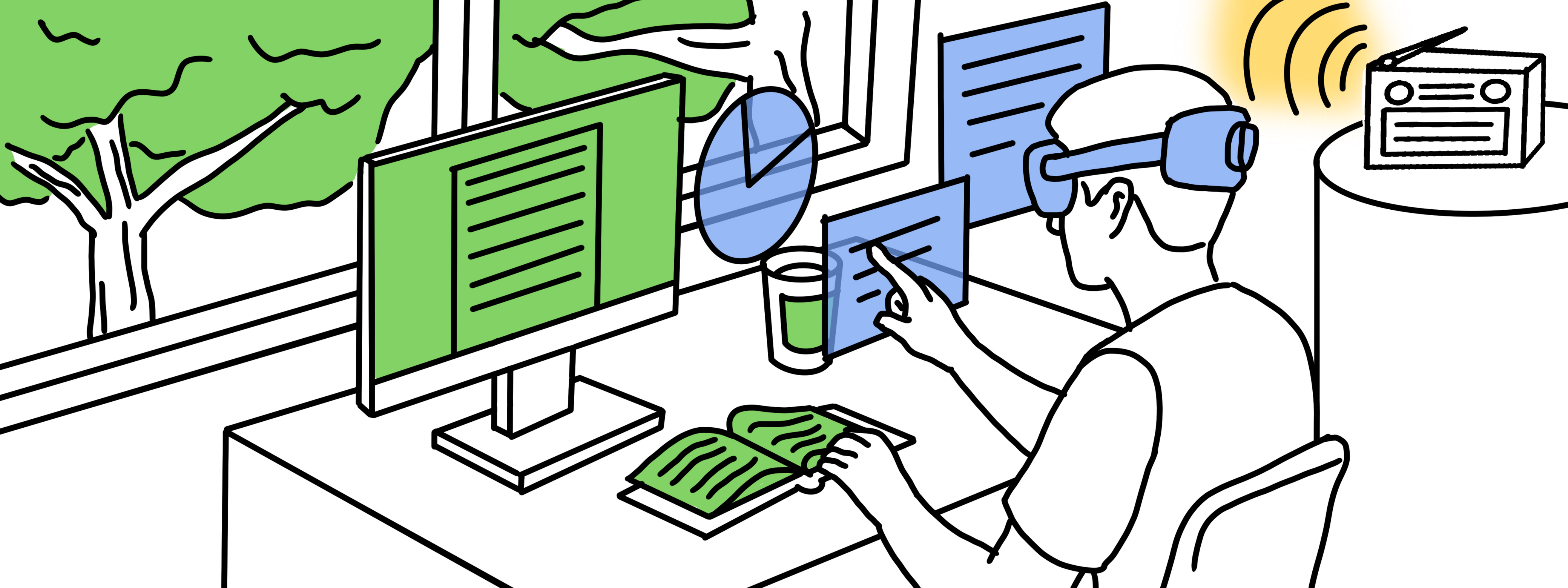}}
    \caption{A user is engaging with virtual content while surrounded by a complex real-world environment, with a radio playing in the background. Green represents the real-world environment, blue represents virtual content, and yellow represents sound. }
    \label{fig:teaser}
    \Description{The illustration depicts a user working at a desk while simultaneously engaging with multiple layers of information in mixed reality (MR). The real-world environment, shown in green, includes a computer monitor, a book, and a view of trees outside the window. Virtual elements, shown in blue, appear as floating screens and interactive content overlaid through the user’s headset. A yellow sound source represents a radio playing in the background. The integration of physical, virtual, and auditory inputs highlights the ways in which users distribute their attention in MR between real-world surroundings, virtual elements, and environmental distractions.}
\end{teaserfigure}

\begin{abstract}
Visual search is a core component of mixed reality (MR) interactions, influenced by the complexities of MR application contexts. In this paper, we investigate how prevalent factors in MR influence visual search performance and spatial regularity memory --- including the physical environment complexity, secondary task presence, virtual content depth and spatial layout configurations. Contrary to prior work, we found that the secondary auditory task did not have a significant main effect on visual search performance, while significantly elevating higher perceived workload measures in all conditions. Complex environments and varied virtual elements depths significantly hinder visual search, but did not significantly increase perceived workload measures. Finally, participants did not explicitly recognize repeated spatial configurations of virtual elements, but performed significantly better when searching repeated spatial configurations, suggesting implicit memory of spatial regularities. Our work presents novel insights on visual search and highlights key considerations when designing MR for different application contexts.
\end{abstract}



\maketitle

\section{Introduction}
As a medium that visually integrates virtual artefacts into physical spaces, the effectiveness of mixed reality (MR) applications is dependent on how efficiently users can visually search for relevant virtual or physical elements within the MR environment. Take for example a scenario where we want to press a virtual MR button while working in a busy office space (see Figure~\ref{fig:teaser}). To do so, we must first locate the button within our visual field. However, the button hovers among other virtual icons scattered at different depths, while the surrounding desks, monitors, and people in the physical world compete for our attention. Other senses may also compete for our attentional resources, such as an announcement from the office telecommunication device. Such seemingly common scenarios illustrate a broader phenomenon in MR use: Locating a target swiftly and accurately amid attentional interference is far from trivial --- as the physical environment's complexity, depth variation of relevant objects, and additional task demands in MR together disrupt attention and hinder search efficiency in ways distinct from traditional displays. 

In daily life, visual search is often disrupted by the presence of physical distractors. In MR, the superimposition of virtual elements and physical objects can reduce the visibility of real objects and divert attention to irrelevant virtual elements~\cite{syiem2020enhancing, maag2023measuring, herbert2017augmented}. Since users interact with MR in diverse and dynamic environments, the interplay of physical and virtual elements, with virtual content often appearing at different depths, introduces additional complexity and makes it particularly challenging to design MR applications that support efficient visual search.

To investigate the sources of visual search interference, studies typically manipulate the properties of physical and virtual objects. Examples include varying the depth~\cite{lee2024visual, mcsorley2001visual, huang2022effects}, size~\cite{krause2017interaction, kia2021effects}, colour~\cite{d1991color, amano2012visual}, lightness~\cite{zhang2022lightness}, and similarity of distractors relative to the target~\cite{feria2012effects, yang2014impact}, as well as adjusting the visibility~\cite{sutton2022look, zhou2021vergence}, density~\cite{trepkowski2019effect} and visual realism~\cite{lee2013effects} of virtual elements, with subsequent effects examined in visual search performance. Based on such designs, many studies attribute visual search interference either to the presence of virtual elements~\cite{chang2014development, mccann1993attentional, wickens2009attentional} or to the features of targets and distractors~\cite{amano2012visual, feria2012effects, sutton2022look}. However, the studies on visual search exhibit certain limitations. First, most studies have been conducted on computer screens~\cite{hornof2004cognitive, chun1998contextual} or on smartphones~\cite{luo2015effects, sarsenbayeva2016situational, sarsenbayeva2018effect}. Research confined to two-dimensional displays can depict the spatial arrangement of objects, but it fails to capture the depth relations that are intrinsic to mixed reality. Second, prior studies primarily focused on isolated factors, such as the complexity of the physical environment~\cite{lee2020effects, wang2020effects} or the spatial distribution of virtual objects~\cite{marek2020contextual, plewan2021visual}. However, in practical usage, visual search in MR will be shaped by the interaction of the physical environment and the virtual elements. Overlooking such interactions limits the applicability of existing conclusions. 

Beyond interference from physical and virtual objects, MR contexts frequently involve dual-task demands. In addition to processing visual prompts and monitoring the surrounding environment, users frequently perform auditory tasks such as listening to music or following spoken instructions. These dual-task demands compel individuals to divide and switch attention within limited cognitive resources, intensifying competition for attentional capacity~\cite{koch2018cognitive, kahneman1973attention, wickens2020processing}. This raises a key question: How the presence of a secondary auditory task influences visual search performance in MR settings that combine different levels of physical complexity and varying depth of virtual objects?

Finally, while interference from multiple sources intensifies attentional competition, predictable spatial configurations of virtual elements, known as spatial regularities, offer compensatory cues that support efficient search. By providing contextual guidance, such regularities reduce search difficulty and enhance information extraction~\cite{chun1998contextual, chun2000contextual, chun2003implicit}. However, variations in environmental complexity, the spatial distribution of virtual elements, and task demands can affect how individuals make use of spatial regularities~\cite{sisk2019mechanisms}. Understanding how spatial regularities are exploited under these conditions is essential for informing the design of MR interfaces that better support efficient visual search.

We conducted a study to disentangle the impacts of physical environment complexity, virtual element depth, and dual-task presence on visual search performance and spatial regularity memory in MR. We used a contextual cueing task (i.e., searching for a target ``T'' among distractors ``L'') to evaluate visual search performance~~\cite{chun1998contextual, chun2000contextual}, and a recognition task (i.e., judging repeated versus novel spatial configurations) to measure memory for spatial regularities~\cite{chun1998contextual, smyth2008awareness, cooper2025multitasking}. Visual search performance was assessed with reaction times and error rates, perceived workload was measured using the NASA-TLX questionnaire. Spatial regularity memory was assessed using participants' reaction times to targets presented in repeated and novel spatial configurations in the contextual cueing task and recognition accuracy in the recognition task.

In contrast to prior work, our results show that the secondary auditory task did not exert a significant main effect on visual search performance. However, it significantly increased perceived workload in all conditions. The dissociation between objective performance and subjective experience highlights the importance of balancing what users perceive with the factors that actually constrain search efficiency. Independently, both complex environments and depth variation impaired search performance. Moreover, introducing virtual element depth cues attenuated the impairments caused by complex physical environments, suggesting that depth can serve as a compensatory design feature when MR applications are deployed in 
complex physical environments. The post-hoc pairwise comparisons revealed that the effects of different combination of factors on visual search varied across different spatial configurations, highlighting the need to consider both for the efficiency of visual search and for the ease with which users can exploit consistent spatial regularities. In addition, although memory for spatial regularities is primarily implicit rather than explicit, repeated spatial configurations yielded significantly shorter reaction times than novel configurations in all conditions. These findings suggest that implicit spatial regularity memory remains robust even when visual search is challenged by other factors.

Our work contributes to better understanding of visual search and spatial regularity memory in MR. By teasing apart the effects of physical environment complexity, virtual element depth, and task type, our findings help identify relevant contextual and design factors for deploying MR in different scenarios. Our discussion further highlights key considerations for designing MR across diverse application contexts.

\section{Related Work}
\subsection{Visual Search}
Visual search has long served as a foundational paradigm for investigating how humans deploy attention and locate targets in complex scenes~\cite{treisman1980feature, palmer2000psychophysics, wolfe1989guided, wolfe2010visual, wolfe2017five, wolfe2020visual,wolfe2021guided}. It is a perceptual task that involves actively scanning a visual scene to locate a particular object or feature among distractors~\cite{wolfe2010visual, wolfe2020visual, hershler2009importance}. Beyond the laboratory, it is an integral part of everyday and professional activities, including finding a friend in a crowd~\cite{juth2005looking, hong2025implicit}, identifying a specific product on a supermarket shelf~\cite{gidlof2017looking, bialkova2020desktop}, and detecting anomalies in medical images~\cite{brams2020focal, drew2013informatics}.

In the context of digital technology such as MR, visual search plays a crucial role in the technology's effective use and design~\cite{rokhsaritalemi2020review, chiossi2024searching, lee2013effects}. However, the novelty of MR, which blends familiar physical environments with potentially unfamiliar and dynamic digital content, modulates how users perform visual search for relevant information~\cite{sonntag2022mixed, stevens2015visual, rappa2022use,li2025weight}. Prior work has shown that technology augmenting the physical world with digital content can have both positive and negative effects on visual search. For instance, \citet{smith2017effects} demonstrated that presenting virtual elements through a Head-Up Display (HUD) in the driver’s field of view improved the efficiency in scanning pseudo-text for a target letter, without producing significant decrements in driving performance. However, changes in visual search performance cannot be attributed solely to spatial distribution of virtual elements~\cite{syiem2021impact}, as the intrinsic characteristics of the virtual elements themselves also exert influence~\cite{syiem2020enhancing}. \citet{chang2014development} reported that when virtual information strongly captures user attention, they focus disproportionately on virtual content, neglecting visual search in the real environment. 

Although virtual elements can influence visual search performance, their co-occurrence with the physical environment in MR makes it challenging to isolate whether observed effects are attributable to the virtual content or to the surrounding physical environment. \citet{li2025effects} explored visual search in a simulated driving task by manipulating the background complexity and the HUD opacity. They found that a higher background complexity impaired search efficiency. Nevertheless, while this study focused on the occlusion effects of virtual elements and environments, the distinct roles of physical and virtual factors in causing visual search impairment remain unexplored. Moreover, variations in visual search performance may arise not only from the complexity of physical or virtual elements, but also by the distinctive properties that are inherent to each. For instance, \citet{kim2025go} reported that, even when the visual appearance of physical and virtual targets was matched, physical targets were detected and discriminated more accurate than virtual ones.

The spatial distribution of virtual objects in MR are shaped by both their relative distances and depth positions. \citet{mcsorley2001visual} examined eye movements during visual search and found that depth cues can supported efficient target detection. However, prior studies on visual search have primarily relied on traditional two-dimensional displays~\cite{hornof2004cognitive, chun1998contextual, luo2015effects, sarsenbayeva2016situational, sarsenbayeva2018effect}, which cannot effectively capture these depth relationships. To overcome the limitations of traditional two-dimensional research, some studies examined visual search in immersive virtual reality (VR)~\cite{beitner2023flipping, ajana2023feature}, augmented reality (AR)~\cite{trepkowski2019effect, van2020drone}, and MR~\cite{perelman2022visual, lee2013effects} environments. For instance, \citet{ward2020immersive} examined how different spatial placements of virtual elements (i.e., list, grid, arc) affect search in VR, reporting faster scanning with the arc layout but higher user preference for list and grid. A key limitation of these studies is that they often focus solely on the role of the features and layouts of the virtual elements, overlooking the influence of the virtual element depth and the physical environment.

In addition, the number of virtual objects in the scene plays a crucial role in visual search. In parallel feature search, when the target differs from distractors by a single basic feature, it can be detected with little or no additional cost as the number of items increases~\cite{thornton2007parallel, mcelree1999temporal}. By contrast, in serial search, reaction times typically increase as the display contains more items~\cite{williams1997patterns, gilden2010serial}. However, visual search is neither purely parallel nor purely serial. Instead, it is a hybrid process that combines parallel guidance with serial inspection of items~\cite{wolfe2021guided}.

Many studies on factors that influence visual search have focused solely on the environment or virtual elements, neglecting the critical role of the task itself. \citet{syiem2021impact} demonstrated that longer reaction times to virtual elements were primarily driven by dual-task contexts rather than by the mere presence of virtual elements. Dual task situations frequently emerge in MR given the need to perform real-world actions, such as walking and avoiding obstacles, while interacting with virtual objects~\cite{abbas2024unveiling, tugtekin2023effect, north2021effects, li2025estimating,li2025trends}. Visual search is sensitive to the cognitive demands associated with dual-task contexts. \citet{jackson2023effects} reported that concurrent visual tasks substantially impair visual search. Even when multitasking involves different sensory modalities, it can still disrupt visual search performance, but such cross-modal combinations caused less interference than two concurrent visual tasks~\cite{jackson2023evaluating}. 

The interaction between physical environments~\cite{bennett2021assessing, lee2019effects}, virtual elements~\cite{kim2022investigating}, and task type~\cite{nian2023effects, solman2011memory, vanrullen2004visual} collectively shape visual search performance, prompting us to inquire how these factors interact and influence visual search in MR. This inquiry is fundamental for understanding visual search, which is essential for ensuring efficient interaction and reducing cognitive load in MR systems.

\subsection{Spatial Regularity}

Spatial regularity refers to the consistent co-occurrence of particular target locations with specific spatial layouts. The resulting enhancement of search efficiency, when observers learn and exploit these regularities, is commonly termed the contextual cueing effect (CCE)~\cite{chun1998contextual, chun2000contextual, chun2003implicit}. For instance, when using Google Chrome, users rapidly locate the `x' button to close tabs because it consistently appears at the right side of each tab. The presence of consistent spatial configurations allows individuals to anticipate target locations, thereby improving search efficiency.

MR environments typically integrate rich environmental backgrounds with complex spatial configurations of virtual elements. This raises the question of whether spatial regularity learning in MR is driven by a user's memory of the presented virtual elements' spatial configurations or by the holistic representations of the MR environment scene. \citet{rosenbaum2013interaction} explored spatial regularity learning by training with predictive scenes and spatial configurations of objects. They observed that retaining only the scene preserved the benefits that spatial regularities provide for visual search, whereas retaining only the spatial configuration eliminated it, suggesting that the scene cue overshadowed the spatial configuration cue. In contrast,~\citet{brooks2010nesting} reported that spatial configurations alone can induce spatial regularity learning with varied background scenes. Such differences in conclusions may stem from overlooking the complexity of the scene, whose variation can influence our memory~\cite{chai2010scene, kyle2025scene}.

In MR, both the planar arrangement of virtual elements and their depth relationships can be varied simultaneously, shaping how users search for and process visual information. \citet{zang2017contextual} examined spatial regularity learning in stereoscopic 3D visual search with items confined to either a near or a far depth plane. They found that swapping the near and far planes preserved the benefit of spatial regularities, whereas swapping the left and right halves of the display disrupted this benefit, suggesting that learning depends primarily on planar inter-item relations rather than depth-defined relations. As virtual elements in MR are typically distributed across multiple depth planes, the ecological validity of this study is limited.

Engaging in multiple tasks can also modulate the formation and expression of learned spatial regularities~\cite{tavera2025role, vicente2022role, annac2013memory}. \citet{cooper2025multitasking} incorporated a secondary tone counting task into the contextual cueing task and reported that multitasking impaired the acquisition and application of contextual cues. However, the impact of multitasking on spatial regularity learning is not limited to attenuation. With limited cognitive resources~\cite{franconeri2013flexible, lieder2018anchoring,lieder2020resource}, occupation of working memory by an additional task can severely impair both the formation and expression of spatial regularity learning, and in some cases eliminate the expression of learned spatial regularities entirely~\cite{chen2019executive}. The multitasking condition will hinder users ability to detect and internalize spatial regularities, thereby constraining their use of such regularities to achieve efficient and accurate visual search. For example, in an MR navigation app, users engaged in auditory notifications may overlook the consistent placement of navigational virtual arrows, forcing them to relocate the arrow each time instead of exploiting the repeated spatial cue to streamline the search.

MR blends the physical world, virtual elements, and task demands into a unified interactive space~\cite{speicher2019mixed, chiossi2024understanding}. However, most studies manipulate only a limited subset of these contextual factors. The studies that have examined the effects of virtual elements on spatial regularity learning~\cite{marek2020contextual} and visual search~\cite{smith2017effects, kim2025go} do not distinguish whether these effects are driven by the elements themselves or by their spatial locations. In addition, overlooking the potential impact of environmental complexity may have been a contributing factor to the inconsistencies reported in prior studies~\cite{chai2010scene, kyle2025scene}. Confounding effects related to the task itself are also a major source of these uncertainties~\cite{syiem2021impact}. Such limitations obscure the effects, leading to theoretical ambiguity and inconsistent findings. By disentangling these factors, we can gain a more precise understanding of the contextual factors that genuinely shape visual search and the memory of spatial regularities in MR.

\section{Method}
Our study aims to to disentangle the impact of virtual element depth, physical environment complexity, and task type on visual search performance and spatial regularities memory. Based on prior evidence that dual-task contexts significantly increase visual search time~\cite{syiem2021impact, jackson2023effects, jackson2023evaluating}, we hypothesize:

\begin{itemize}
  \item \textbf{[H1]} In MR, dual task will significantly impair visual search performance.
\end{itemize}

Visual search performance in MR is also shaped by the visual properties of the surrounding scene. Prior work on perceptual load and scene clutter indicates that visually complex backgrounds are associated with slower target detection~\cite{li2025effects, lee2020effects}. Thus, we hypothesize:

\begin{itemize}
  \item \textbf{[H2]} In MR, performing visual search in a visually complex physical environment will significantly impair visual search performance.
\end{itemize}

Additionally, prior studies on 3D visual search suggests that virtual element depth can act as a guiding attribute for attention in visual search~\cite{wolfe2017five}, particularly when targets are associated with a distinct or cued depth plane~\cite{zou2022top}. In MR, virtual element depth cues can segment the display into separable layers. Consequently, we hypothesise:

\begin{itemize}
  \item \textbf{[H3]} In MR, presenting virtual elements at different depth planes will support visual search.
\end{itemize}

Furthermore, the acquisition of spatial regularities during visual search has been shown to vary in response to dual-task demands, variable depth cues, and environmental complexity. Under dual-task conditions, attentional resources are divided across concurrent tasks, and spatial regularity learning is reduced or constrained~\cite{cooper2025multitasking, chen2019executive}. In contrast, prior evidence suggests that spatial regularity learning can still be observed under changes in virtual element depth~\cite{zang2017contextual} and in dynamic environments~\cite{chun1999top}. Therefore, we hypothesize:

\begin{itemize}
  \item \textbf{[H4]} In MR, spatial regularity learning is conditional rather than universal: the formation of spatial regularity learning will be sensitive to the presence of a dual task, but remain robust to variations in virtual element depth and physical environment complexity.
\end{itemize}

H1–H3 hypothesize main effects for task type, physical environment, and virtual depth. In addition to these main effects, we also anticipated interaction effects, such that the influence of each factor would vary as a function of the others. H4 focuses on the conditions under which spatial regularity learning can be formed and expressed. To test these hypotheses, we conducted a user study using a custom MR application that allowed us to systematically control the relevant experimental factors and measure both visual search performance and spatial regularity memory. Additionally, we employed the NASA-TLX questionnaire to gain insights about users' perceived workload. All study materials and procedures were approved by our institution’s ethical review panel.

\begin{figure*}[tbp]
    \centering
    \includegraphics[width=0.95\textwidth, height=4.4cm]{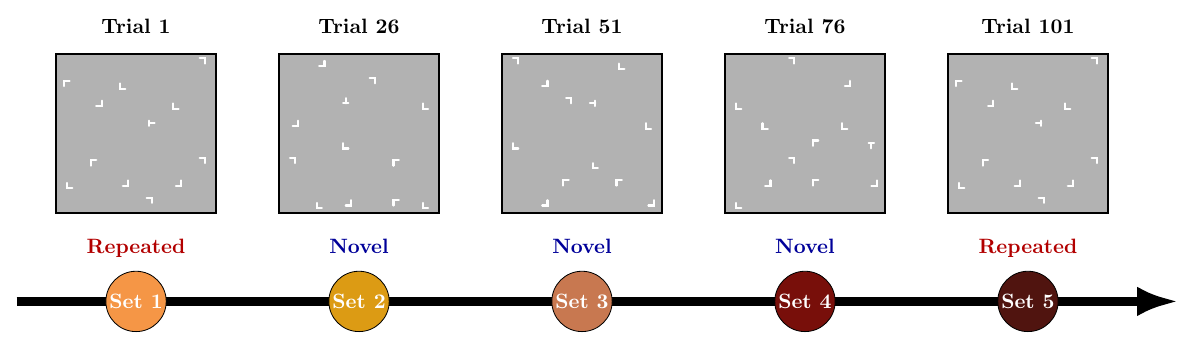}
    \caption{Overview of the contextual cueing task procedure. Each trial displayed a visual search array with one virtual target ``T'' among eleven virtual distractor ``L''. Repeated spatial configurations, exemplified at Trial 1 and Trial 101, preserved the same spatial arrangement of distractors, with only the target orientation varied across repetitions. Novel spatial configurations, exemplified at Trial 26, Trial 51, and Trial 76, were newly generated for each trial. The timeline illustrates how repeated and novel spatial configurations appeared across trials. In the experiment, the trial order of repeated and novel spatial configurations was randomized.}
    \Description{The figure illustrates the procedure of the contextual cueing task used in the experiment. In each trial, participants were presented with a visual search array containing one target item (a rotated “T”) embedded among eleven distractor items (“L”s). The task required participants to detect and respond to the orientation of the target as quickly and accurately as possible. Two types of spatial configurations were employed. Repeated spatial configurations, as exemplified in Trial 1 and Trial 101, preserved the same spatial arrangement of distractors across multiple trials, with only the orientation of the target varying across repetitions. Novel spatial configurations, such as those shown in Trial 26, Trial 51, and Trial 76, were newly generated for each presentation, ensuring that participants could not rely on prior exposure. The timeline at the bottom of the figure depicts the distribution of repeated and novel configurations across five experimental sets, each comprising 25 trials. Although repeated and novel configurations are shown in alternation for illustration, in the actual experiment their trial order was randomized. This paradigm allows researchers to examine whether participants learn and exploit repeated spatial regularities to facilitate visual search performance over time.}
    \label{fig:contextual cueing task}  
\end{figure*}

\subsection{Experimental Task}
The experimental tasks were administered in two sequential phases. The first phase consisted of a contextual cueing task~\cite{chun1998contextual, chun2000contextual} that involved searching for a virtual target among distractors in 125 trials (\(5 \ sets \times 25 \ trials \) featuring repeated or novel spatial configurations of the virtual elements. The second phase employed a recognition task~\cite{smyth2008awareness, cooper2025multitasking} that assessed explicit memory of spatial regularities in 4 trials. The task required participants to determine if a given spatial configuration of virtual elements appeared repeatedly during the first phase. We detail these two task in the following subsections.

\subsubsection{Contextual Cueing Task}
The task was designed based on the contextual cueing paradigm experiment~\cite{chun1998contextual, chun2000contextual}, in which participants performed a visual search while learning repeated spatial configurations. This paradigm allowed us to assess users' visual search performance by measuring their reaction times in locating a virtual target. Additionally, by comparing reaction times to repeated versus novel spatial configurations, we could also assess whether participants memorized spatial regularities.

The virtual elements were presented on an invisible \(6 \times 8\) grid, with each cell measuring 35 cm. The \(6 \times 8\) grid provides 48 potential stimulus locations, offering sufficient variability for constructing repeated and novel spatial configurations while being consistent with the set sizes typically used in contextual cueing paradigms~\cite{chun1998contextual, tseng2013rewarding, zhao2021contextual}. The cell size of 35 cm ensured adequate spacing between virtual elements to avoid visual crowding and prevent overlap between neighboring items. On each trial, one virtual target letter ``T'' and eleven virtual distractor letters ``L'' were randomly placed within the grid, with at most one stimulus per cell. This set size of 12 items aligns with standards commonly employed in contextual cueing paradigms~\cite{chun1998contextual, thomas2018limits, zhao2021contextual}. Both ``T'' and ``L'' were randomly rotated to one of four orientations (0$^\circ$, 90$^\circ$, 180$^\circ$ and 270$^\circ$). For the virtual target ``T'', the four possible orientations were directly mapped to the arrow keys on the keyboard: with the down-arrow key relating to a T without any rotation, the left-arrow key relating to a T with 90-degree clockwise rotation, the up-arrow key relating to a T with 180-degree rotation, and the right-arrow key relating to a T with 270-degree clockwise rotation. The orientation discrimination of the virtual target “T” was included to reduce guessing responses~\cite{chun1998contextual}. Although this manipulation may increase cognitive load, it was applied uniformly so that differences in performance and perceived workload can be attributed to the experimental factors rather than to the orientation-discrimination requirement itself. Each trial was separated by an interval of 500 ms during which no virtual elements were presented, preventing visual persistence from the previous trial~\cite{makovski2016context, jiang2019contextual}.

We used two types of spatial configuration: repeated spatial configurations and novel spatial configurations. Repeated and novel spatial configurations were used to assess whether participants acquired memory for spatial regularities, which would result in faster search times for repeated spatial configurations compared to novel spatial configurations~\cite{chun1998contextual, chun2000contextual}. In repeated spatial configurations, apart from the orientation of the virtual target ``T'', the spatial position of the target and the positions and orientations of all distractor ``L'' were kept constant across trials. In novel spatial configurations, each trial presented a newly generated random spatial configuration. This manipulation ensured that the repeated benefit could not be attributed to low-level feature repetition (e.g., an identical target orientation), but rather to the preserved spatial arrangement of the target and distractors~\cite{treisman1980feature}.

The task was structured into five sets of twenty-five trials. This number was implemented to ensure that the session duration remained manageable and to minimize the risk of participant fatigue. Within each set, five repeated spatial configuration trials and twenty novel spatial configuration trials were presented in randomized order. This proportion of repeated to novel spatial configuration trials is consistent with ratios employed in previous CCE experiments~\cite{hatori2024modeling, zinchenko2018predictive}. To prevent participants from simply recognizing the repeated spatial configurations from the immediately preceding trial, at least two novel spatial configurations were inserted between any two repeated spatial configurations. Between sets, a three-second interval was provided for rest and attentional reorientation. Figure~\ref{fig:contextual cueing task} shows an illustrative example of the contextual cueing task procedure.

\begin{figure*}[tbp]
    \centering
    \begin{subfigure}{0.48\textwidth}
        \centering
        \includegraphics[width=\linewidth, height=5.3cm]{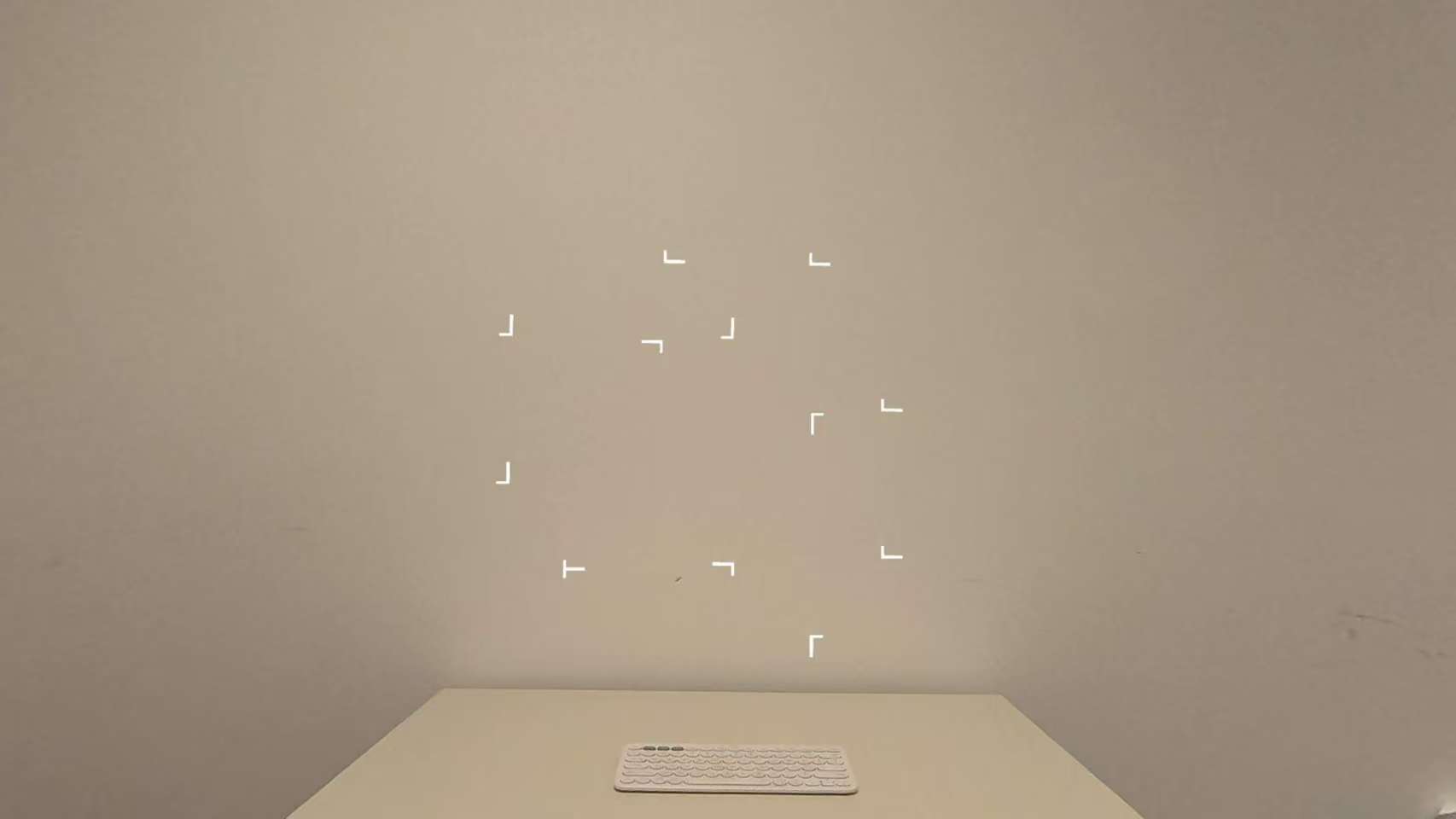}
        \label{fig:simple environment}
    \end{subfigure}
    \hfill
    \begin{subfigure}{0.48\textwidth}
        \centering
        \includegraphics[width=\linewidth, height=5.3cm]{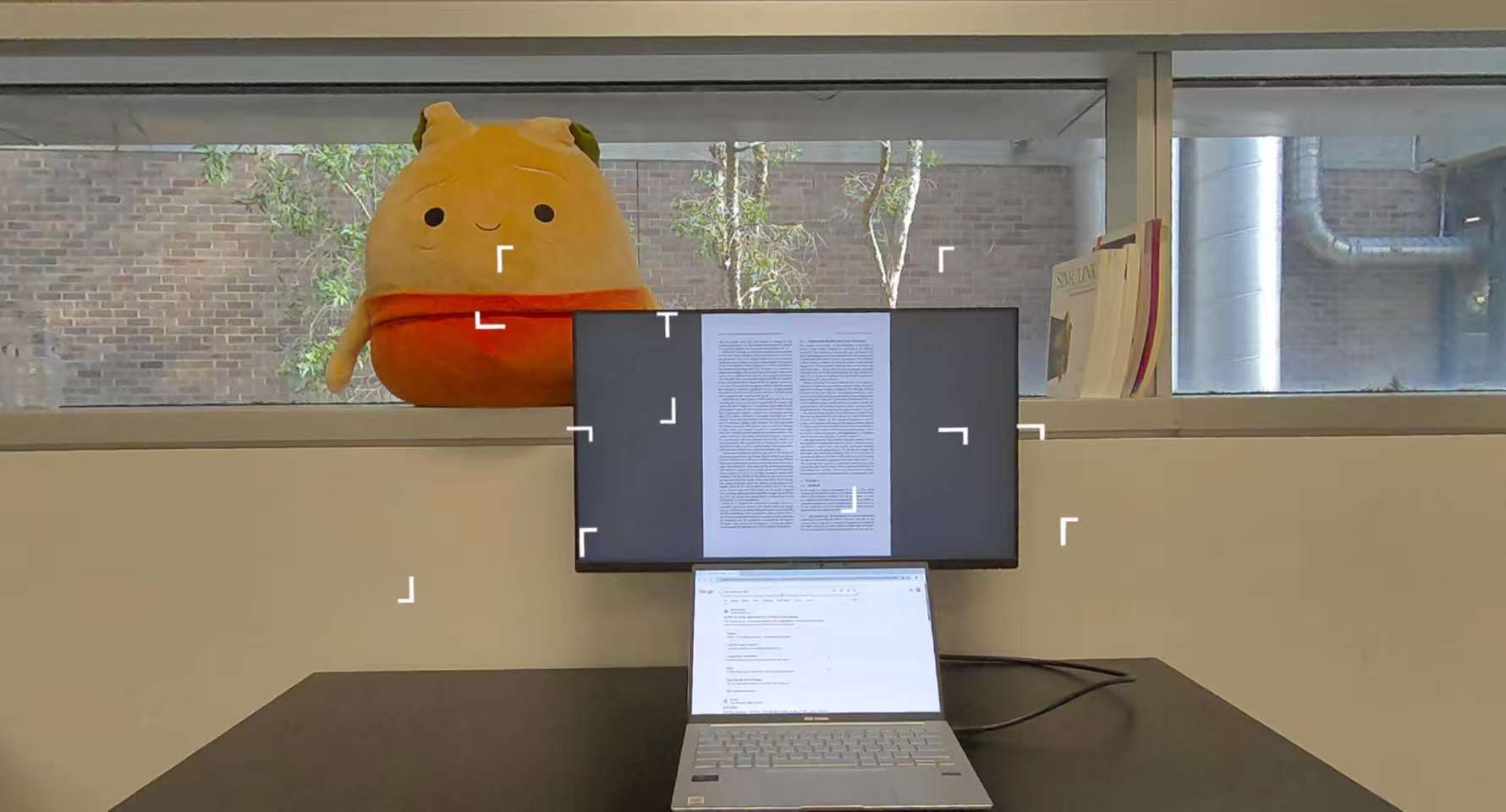}
        \label{fig:complex environment}
    \end{subfigure}
    \caption{Experimental environments. Left: The simple environment with virtual elements that were presented at the same depth, consisting of a white desk against a white wall, with only a white keyboard on the desk. Right: The complex environment with virtual elements that were presented at different depths, incorporating a monitor, laptop, plush toy, and outdoor background scene visible through a window. While our MR system varied the depth at which virtual elements appeared in the different-depth condition, such variations are difficult to illustrate with 2D images. In 2D, depth manipulations may appear as changes in stimulus size. However, changing size of virtual elements does not have the same effects as changing the depth, as depth variations require users to refocus their eyes to different depth planes~\cite{deng2021towards, lee2024visual, tsirlin2016size}.}
    \Description{The images show two types of experimental environments used in the study. The left panel depicts the simple environment, consisting of a white desk against a plain wall with only a white keyboard placed on the desk. Virtual stimuli, represented by the scattered “L” and “T” shapes, were presented at the same depth in this setting. The right panel illustrates the complex environment, which included multiple physical objects such as a monitor, laptop, and plush toy, as well as an outdoor view through the window. In this condition, virtual stimuli were presented at different depths. Although depth was manipulated in the MR system, these variations are not easily conveyed in static 2D images. In two-dimensional representations, depth changes may resemble differences in stimulus size. However, in MR, varying depth requires users to refocus their eyes to distinct depth planes, making it a qualitatively different experience.}
    \label{fig:physical environment}
\end{figure*}

\subsubsection{Recognition Task}
In the contextual cueing task, faster responses to repeated spatial configurations compared to novel ones indicate more efficient visual search facilitated by memorized spatial regularities. However, such behavioral facilitation alone cannot establish whether participants possessed explicit memory of the repeated spatial configurations~\cite{chun1998contextual, cooper2025multitasking}. To clarify this distinction, we included a recognition task~\cite{chun1998contextual, cooper2025multitasking, smyth2008awareness} that directly assessed explicit memory, allowing us to determine if participants were aware of repeating spatial configurations during the contextual cueing task. Participants were asked to perform the recognition task three seconds following completion of the contextual cueing task. In this task, four spatial configurations were sequentially presented in random order, consisting of the one repeated spatial configuration from the contextual cueing task and three newly generated spatial configurations. By introducing three novel spatial configurations as distractors, the recognition task reduced the likelihood of false positives arising from random guessing. For each spatial configuration, participants were asked to judge whether it was a repeated or a novel spatial configuration.

\subsection{Experimental Design} 
We conducted a \(2 \times 2 \times 2 \) within-subject experiment with three independent variables: physical environment (simple environment vs. complex environment) \(\times\) virtual object depth (same depth vs. different depth) \(\times\) task type (single task vs. dual task), resulting in eight experimental conditions in total. Within each condition, repeated and novel spatial configurations were newly generated and did not reappear in other conditions.

\subsubsection{Independent Variables} Details of the independent variables used in our experiment are presented below:

\textit{Physical environment complexity.} We designed two levels of physical environmental complexity, a simple environment and a complex environment, to simulate how users engage with MR in two common contexts. The simple environment condition was used to approximate MR use in contexts with reduced visual clutter. In this condition, a plain white wall was positioned directly in front of the participant's field of view, providing a simple visual background with minimal distractions. To further prevent potential color effects on visual search, all equipment, including the keyboard and desk, was set uniformly to white (see Figure~\ref{fig:physical environment} left). In the complex environment, participants were seated at an office desk equipped with a laptop and a monitor, simulating the context of a professional work setting in MR. To reflect the abundance of objects and vivid colors commonly present in real work environments, the experimental environment featured a black table, with toys and books positioned behind the monitor, and a view of the outdoor scenery through the window to increase visual contrast and create a more realistic context (see Figure~\ref{fig:physical environment} right).

\textit{Virtual element depth.} We manipulated the placement of virtual objects under two depth conditions (same depth vs. different depth) to investigate how spatial positioning of virtual elements influences visual search and the memory of spatial regularities. For the same depth condition, both virtual elements ``L'' and ``T'' were presented at 350 cm from the participant on a single depth plane. For the different depth condition, each stimulus was presented at a randomly assigned depth plane between 300 and 450 cm. These depth values were selected guided by previous studies on human depth perception of virtual objects~\cite{waller2008correcting, diaz2017designing}. For repeated spatial configurations, depth-plane assignments were fixed across occurrences within each condition. For novel spatial configurations, depth planes were newly randomized on each trial.

\textit{Task type.} We implemented two task conditions: a single-task condition and a dual-task condition. The single-task condition represented a pure visual search scenario. The dual-task condition incorporated a concurrent auditory task to simulate dual-task situations that users may often face when engaging with MR. For the single-task condition, participants completed the contextual cueing task followed by the recognition task. For the dual-task condition, participants performed the contextual cueing task while concurrently monitoring auditory tones~\cite{cooper2025multitasking, vroomen2000sound, matusz2011multisensory}. Two pure sine tones were used: a low tone at 440 Hz corresponding to the musical note A4 and a high tone at 880 Hz corresponding to the musical note A5, which form a standard octave pair. This pair is easily distinguishable and has been widely adopted in auditory research~\cite{pazdera2025pitch, ono2018modality}. Each tone lasted 250 ms, with an interval of one second between tones. Participants counted the number of high tones within each set and ignored low tones. At the end of each set, they entered the total number of high tones via the keyboard. During the recognition task, the procedures were the same in both conditions, and the participants completed only the recognition responses.

\subsubsection{Dependable Variables}

After completing the experimental tasks, we collected four categories of data to evaluate task performance and user perceptions. The details of each dependent variable are described below:
\textit{Reaction times.} In the contextual cueing task, reaction times served as a direct indicator of search efficiency. We defined reaction times separately for correct and incorrect responses. We defined correct responses as pressing the arrow key matching the target’s orientation, and incorrect responses as any other keypress. We recorded reaction time on each trial. For trials with correct responses, reaction time was the interval between stimulus onset and the correct keypress. For trials with an incorrect initial response, we recorded only the number of erroneous keypresses, until a correct response, and the corresponding reaction times were excluded from data analyses. We calculated the difference in reaction times between repeated and novel spatial configurations within each condition to assess spatial regularity memory.

\textit{Keypress accuracy.} We defined correct responses as keypresses that matched the target’s orientation, and we calculated the accuracy for each condition as the proportion of correct responses relative to the total number of trials completed by a given participant. Combined with reaction times, this measure offered a more comprehensive assessment of visual search performance.

\textit{Recognition error rate.} We assessed explicit memory for spatial regularities in the recognition task. We considered the error rates in distinguishing repeated from novel spatial configurations as index of explicit memory of spatial layouts, with lower error rates reflecting stronger recognition.

\textit{Perceived workload.} We assessed perceived workload with the NASA-TLX questionnaire, which includes six subscales: mental demand, physical demand, temporal demand, performance, effort, and frustration. We analyze the individual NASA-TLX subscale scores to provide an estimate of perceived workload under eight experimental conditions (\(2 \ physical \ envrionment \ complexity \times 2 \ virtual \ element \ depth \times 2 \ task types\)).

\subsection{Experimental Setup}
The experiment was implemented using a custom MR application in Unity 6000.1.0f1 and deployed on a Meta Quest 3 headset. The system integrated the physical environment with overlaid virtual elements (see Figure~\ref{fig:physical environment}). An invisible \(6 \times 8\) grid was anchored at the center of the participant's starting position when the task began. Virtual elements were rendered stereoscopically within the headset. A keyboard connected to the headset was used for orientation responses, and the application automatically logged reaction times, response key and task events. Each trial proceeded only after the participant pressed the correct key corresponding to the orientation of the virtual target ``T''. We conducted the study in a quiet room (\(5 m \times 6 m\)) with only one participant and one experimenter present.

\subsection{Participants}
We recruited twenty-four participants (M = 12, F = 12) aged between 21 and 29 (M = 24.38, SD = 2.16) for the study. This sample size is consistent with standards in HCI research~\cite{caine2016local}. All participants had normal or corrected-to-normal vision.

\subsection{Procedure}
Our study followed a within-subject design in which each participant completed all eight conditions. To mitigate sequence effects, the order of conditions was counterbalanced using a Latin square.

Upon participants' arrival at the laboratory, we introduced the study procedures and answered their questions. After confirming their understanding and willingness to participate, participants signed the consent form. Subsequently, participants completed a demographic questionnaire collecting data on their age, gender, and prior experience with MR.

To ensure that participants understood the task, they completed a tutorial which introduced the task procedures and allowed them to practice before the formal experiment. In the tutorial, participants completed one practice set under the single task condition, which comprised twenty-five contextual cueing task trials and one recognition task trial, followed by an equivalent practice set under the dual task condition. The training phase was intended to familiarize participants with the procedures and was excluded from the statistical analyses. If participants remained uncertain about the task, we provided additional explanations and extra practice trials. All spatial configurations of target and distractors in the tutorial practice trials were excluded from the main experiment.

During the formal experiment phase, participants completed the contextual cueing task followed by the recognition task under all eight conditions. Upon completing both tasks within a condition, participants filled out the NASA-TLX questionnaire to assess the subjective workload. Flexible breaks were offered between conditions, and participants proceeded once they felt adequately rested.

\begin{figure*}[tbp]
    \centering
    \includegraphics[width=\textwidth]{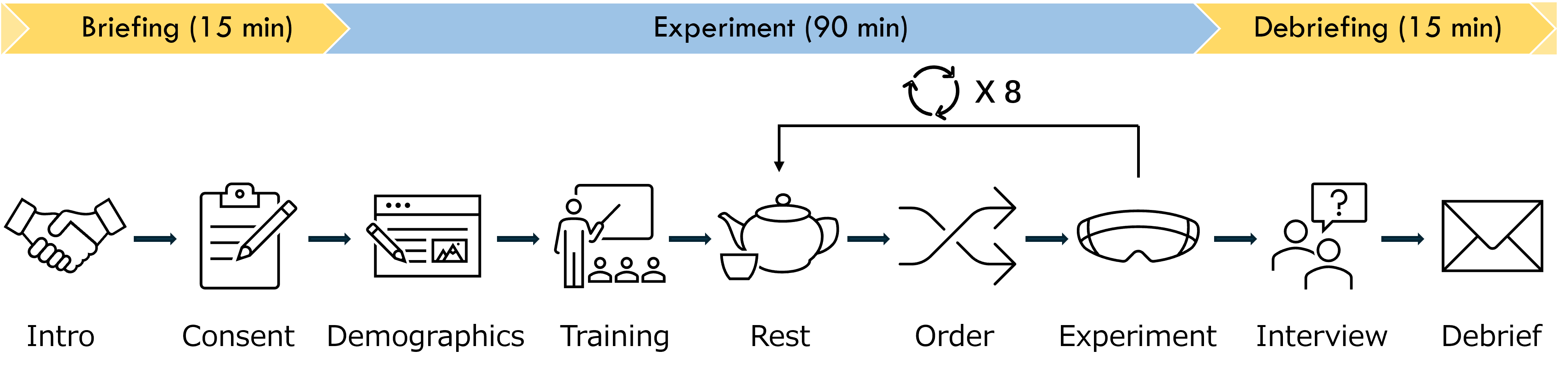}
    \caption{Overview of the experimental procedure, consisting of a 15-minute briefing session, a 90-minute main experiment composed of eight randomized conditions separated by rests, and a 15-minute debriefing phase with interview and discussion.}
    \Description{The figure illustrates the overall structure of the experimental procedure. The study began with a 15-minute briefing session, during which participants were welcomed, provided consent, completed demographic questionnaires, and received training. This was followed by the main experimental phase, lasting approximately 90 minutes. In this phase, participants completed eight conditions, with the order counterbalanced across participants and short breaks inserted between conditions to reduce fatigue. Finally, a 15-minute debriefing phase was conducted, which included a semi-structured interview and a final discussion to gather participants’ feedback and reflections.}
    \label{fig:experimental procedure}
\end{figure*}

At the end of the study, we conducted a semi-structured interview to gain supplementary insights into their experience
during the study. The total duration of the study was 120 minutes per participant. The complete process is illustrated in Figure~\ref{fig:experimental procedure}.

\subsection{Data Analysis}
We used Linear Mixed-Effects Models (LMMs), Generalized Linear Mixed-Effects Models (GLMMs), and Cumulative Link Mixed Models (CLMMs) to model the effects of experimental factors on the dependent variables. We applied LMMs to reaction times, as they allow continuous responses to be modelled while accounting for both fixed effects of experimental factors and random effects associated with participants. We used GLMMs with a logit link function to analyze accuracy, which followed a binary distribution, enabling us to model accuracy while incorporating random intercepts for participants. We employed CLMMs to examine the subjective workload ratings of each subscale of the NASA-TLX questionnaire, as they are specifically designed for the ordinal data analysis~\cite{christensen2018cumulative}. To isolate the effects of our independent variables on the different spatial configurations, we also conducted Tukey-adjusted pairwise comparisons between conditions for the combined and individual spatial configurations. These comparisons allowed us to assess whether the effects of the independent variables were robust across configurations or emerged only under specific spatial configurations. All pairwise comparisons were based on the estimated marginal means (EMMs) derived from the respective LMM or GLMM model fits~\cite{searle1980population, lenth2023emmeans}.

\section{Results}
We collected a total of 24,728 responses (\(8 \ conditions \times 5 \ sets \times 25 \ trials \times 24 \ participants  + incorrect \ responses\)) in the contextual cueing task and 768 (\(24 \ participants \times 8 \ conditions \times 4 \ trials\)) in the recognition task. Responses with reaction times shorter than 100 ms were treated as accidental key presses and excluded~\cite{luce1991response}, leaving 24,727 records for the contextual task. We retained all recognition task responses as they met the inclusion criteria.

\subsection{Visual Search Performance}
\subsubsection{Linear Mixed-Effects Model for Reaction Times}
We analyzed log-transformed reaction times using LMMs, following the common approach in cognitive psychology and HCI to reduce skewness and approximate normality~\cite{baayen2008mixed, baayen2010analyzing}. The fixed effects comprised of physical environment complexity, virtual element depth, task type and spatial configurations, together with their interactions. Random intercepts were included to capture individual variability across participants. The reference condition was the combination of the simple environment, same-depth virtual elements, a single-task setting, and novel spatial configurations. The intercept represents the mean log reaction times in the reference condition. 

Table~\ref{tab:Response Time LMM} presents the regression coefficients ($B$) from the LMM fit, along with their standardized regression coefficients ($\beta$), standard errors (SE), \textit{t}-values, and \textit{p}-values. Positive $B$ values correspond to longer reaction times, suggesting increased difficulty in locating the target and negative $B$ values correspond to shorter reaction times, reflecting more efficient target search relative to the reference condition. $\beta$ denotes the effect size estimated by the LMM~\cite{ben2020effectsize}. The variance of the random intercept was 0.023, with a standard deviation of 0.15. The residual variance was 0.232, with a standard deviation of 0.48. We found that reaction times were significantly longer in the complex environment relative to the simple environment ($B = 0.157$, $\beta = 0.305$, SE = 0.014, \textit{t}(23971) = 11.138, \textit{p} = 9.65E$-29$). We observed a significant increase in reaction time for virtual elements presented at different depths ($B = 0.047$, $\beta = 0.091$, SE = 0.014, \textit{t}(23971) = 3.333, \textit{p} = 8.6E$-4$). We found that repeated spatial configurations reduced reaction time ($B = -0.218$, $\beta = -0.425$, SE = 0.022, \textit{t}(23971) = $-9.824$, \textit{p} = 9.79E$-23$), demonstrating a facilitation effect of repetition. Contrary to prior work~\cite{richard2002effect, gherri2011active}, we find that the main effects of dual-task presence was not significant.

When the complex environment was combined with virtual elements presented at different depths, the coefficient was negative and significant ($B = -0.056$, $\beta = -0.110$, SE = 0.02, \textit{t}(23971) = $-2.838$, \textit{p} = 4.54E$-3$), indicating a sub-additive cost, whereby the combined increase in reaction times was smaller than the sum of the increases produced by the complex environment and the different-depth condition individually. Additionally, we find that the repetition benefit was reduced in the complex environment ($B = 0.063$, $\beta = -0.122$, SE = 0.032, \textit{t}(23971) = 1.985, \textit{p} = 4.71E$-2$) and when virtual elements were presented at different depths ($B = 0.063$, $\beta = -0.123$, SE = 0.031, \textit{t}(23971) = 2.008, \textit{p} = 4.47E$-2$). No other two-way interactions were statistically significant. For the higher-order interactions, a significant three-way interaction emerged when participants encountered repeated spatial configurations in a complex environment with virtual elements that were presented at different depths ($B = -0.138$, $\beta = -0.268$, SE = 0.045, \textit{t}(23971) = $-3.091$, \textit{p} = 2.0E$-3$). The remaining three-way interaction and the four-way interaction were not significant.

\begin{table*}[t]
\centering
\caption{Results of the linear mixed-effects model predicting log reaction time (ms) from environmental complexity, virtual depth, and task type, with random intercepts for participants. The results showed that both complex environments and virtual elements presented at different depths significantly increased response times, while repeated spatial configurations significantly facilitated faster responses. The significant two-way interactions indicated sub-additive joint effects of the complex environment and different-depth virtual elements, and a reduced repeated-configuration benefit under complex-environment and different-depth conditions. A significant three-way interaction was observed among environment complexity, virtual depth, and spatial configuration type.}
\label{tab:Response Time LMM}
\begin{tabular*}{\textwidth}{@{\extracolsep{\fill}}lccccc@{}}
\toprule
\textbf{Parameter} & \textbf{$B$} & \textbf{$\beta$} & \textbf{Std. Error} & \textbf{t(23971)} & \textbf{p-value}  \\
\midrule
(Intercept) & 7.175 & -0.145 & 0.032 & 222.915 & 2.62E-46***  \\
Complex Environment & 0.157 & 0.305 & 0.014 & 11.138 & 9.65E-29***  \\
Different Depth & 0.047 & 0.091 & 0.014 & 3.333 & 8.6E-4*** \\
Dual Task & 0.013 & 0.025 & 0.014 & 0.934 & 3.5E-1 \\
Repeated Configuration & -0.218 & -0.425 & 0.022 & -9.824 & 9.79E-23*** \\
Complex Environment × Different Depth & -0.056 & -0.110 & 0.020 & -2.838 & 4.54E-3** \\
Complex Environment × Dual Task & 0.016 & 0.030 & 0.020 & 0.786 & 4.32E-1 \\
Different Depth × Dual Task & 0.009 & 0.017 & 0.020 & 0.445 & 6.56E-1 \\
Complex Environment × Repeated Configuration & 0.063 & 0.122 & 0.032 & 1.985 & 4.71E-2* \\
Different Depth × Repeated Configuration & 0.063 & 0.123 & 0.031 & 2.008 & 4.47E-2* \\
Dual Task × Repeated Configuration & 0.029 & 0.056 & 0.031 & 0.912 & 3.62E-1 \\
Complex Environment × Different Depth × Dual Task & 0.044 & 0.085 & 0.028 & 1.559 & 1.19E-1 \\
Complex Environment × Different Depth × Repeated Configuration & -0.138 & -0.268 & 0.045 & -3.091 & 2.00E-3** \\
Complex Environment × Dual Task × Repeated Configuration & -0.011 & -0.021 & 0.045 & -0.243 & 8.08E-1 \\
Different Depth × Dual Task × Repeated Configuration & 0.062 & 0.120 & 0.044 & 1.400 & 1.63E-1 \\
Complex Environment × Different Depth × Dual Task × Repeated Configuration & 0.033 & 0.065 & 0.063 & 0.530 & 5.96E-1 \\
\bottomrule
\end{tabular*}
\begin{center}
\footnotesize
\textbf{Note.} B denotes unstandardized regression coefficients, and $\beta$ denotes standardized regression coefficients.

Significance: * \textit{p} < .05, ** \textit{p} < .01, *** \textit{p} < .001.
\end{center}
\end{table*}

\subsubsection{Pairwise Comparisons of Reaction Times}
We first analysed the main effect of each factor on reaction times, using model-based EMMs from LMM computed across combined spatial configurations. For task type, dual-task reaction times did not differ significantly from single-task reaction times only in the simple environment with virtual elements at the same depth. In all other combinations of physical environment complexity and virtual element depth, the dual-task condition yielded significantly longer reaction times. For virtual depth, reaction times differed significantly between same-depth and different-depth presentations across all combinations of task type and physical environmental complexity. For physical environmental complexity, reaction times in the complex environment were significantly longer than in the simple environment across all combinations of task type and virtual depth. The details of the pairwise comparisons between combined spatial configurations are provided in Table~\ref{tab:RT Emmeans All} of Appendix~\ref{appendix:pairwise_comparison}.

Figure~\ref{fig:Reaction Time} illustrates mean reaction times with their standard error across the eight experimental conditions. We can see that mean reaction times were markedly slower in the complex environment compared to the simple environment. With respect to virtual element depth, mean reaction times were generally longer when virtual elements were presented at different depths compared to the same depth, except in the single-task condition of the complex environment, where different depth was marginally faster. While the difference between dual-task and single-task conditions was relatively modest, a substantial increase in mean reaction times for dual-task compared to single-task was observed in the complex environment with different depth. In general, the shortest mean reaction times occurred in the simple environment under single task with same depth, whereas the longest mean reaction times were observed in the complex environment under dual task with different depth.

\begin{figure*}[tbp]
  \centering
  \includegraphics[width=0.97\textwidth]{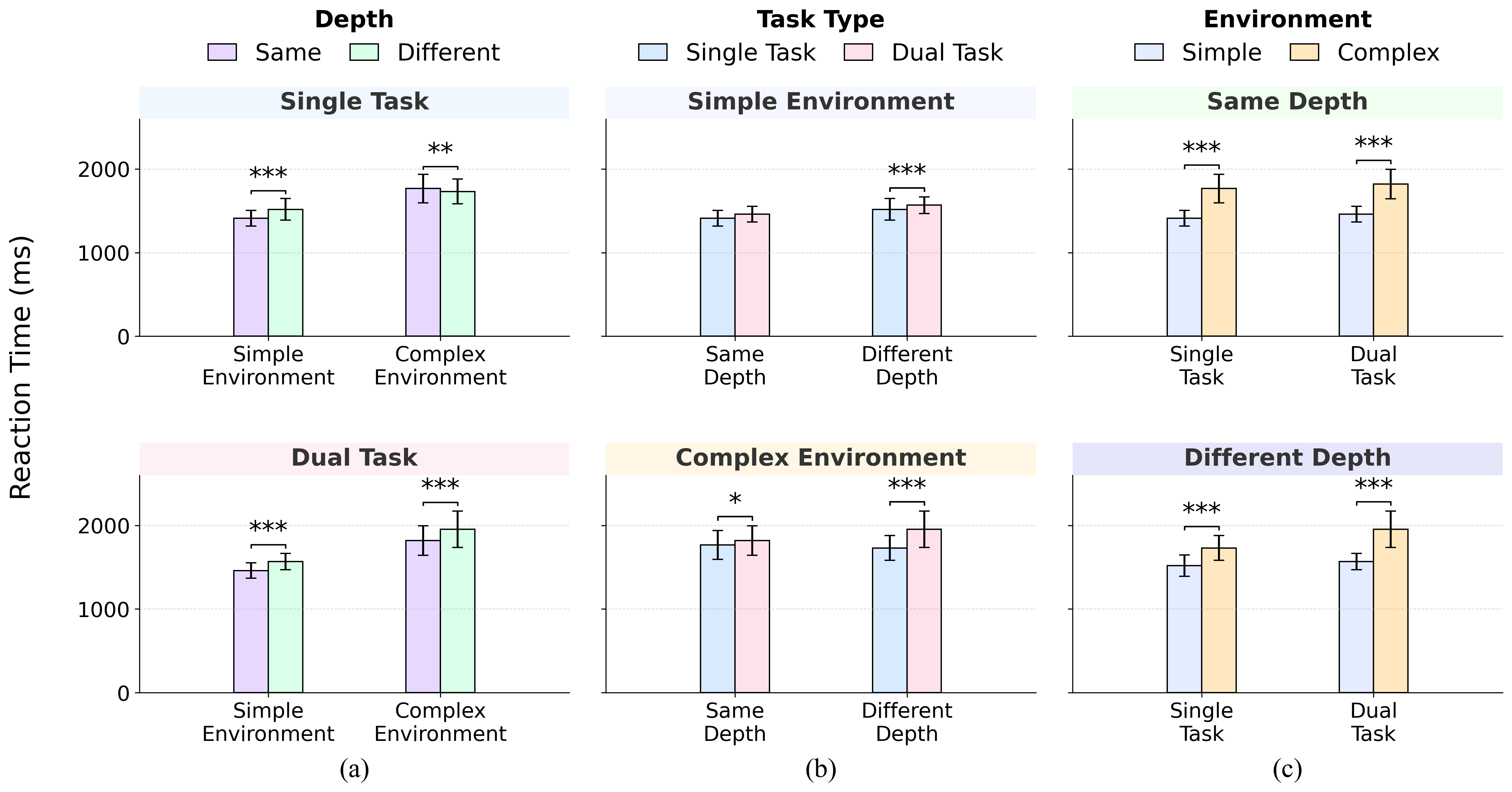}
  \caption{Bar plots for the mean reaction times (ms) of correct trials across three experimental factors: (a) task type, (b) virtual element depth, and (c) physical environment complexity. Statistical significance was assessed using pairwise comparisons of estimated marginal means with Tukey-adjusted p-values. Asterisks denote significant pairwise comparisons (* \textit{p} < .05, ** \textit{p} < .01, *** \textit{p} < .001).}
  \Description{This figure shows bar plots of the mean reaction times to our contextual cueing task. Reaction time was measured in milliseconds from the onset of the target to the participant's correct response, with incorrect trials excluded from the averages. The bar plots show the mean reaction times across three experimental factors: (a) task type, (b) virtual element depth, and (c) physical environment complexity. Asterisks denote significant pairwise comparisons (* \textit{p} < .05, ** \textit{p} < .01, *** \textit{p} < .001). All pairwise comparisons were significant except the task-type comparison between the single-task and dual-task conditions under the simple-environment, same-depth condition.}
  \label{fig:Reaction Time}
\end{figure*}

We then conducted pairwise comparisons restricted to novel spatial configurations. Dual-task reaction times were significantly longer than single-task reaction times in the complex environment for both the same-depth condition and the different-depth condition. The contrast between the same-depth condition and the different-depth condition was non-significant only in the complex environment under single-task performance and significant in all other combinations of task type and physical environment complexity. Finally, reaction times were significantly longer in the complex environment than in the simple environment across all combinations of task type and virtual depth. The details of the pairwise comparisons between novel spatial configurations are provided in Table~\ref{tab:RT_Emmeans_Novel} of Appendix~\ref{appendix:pairwise_comparison}.

Finally, we conducted pairwise comparisons for repeated spatial configurations. We found that the dual-task condition yielded significantly longer reaction times than the single-task condition in the different-depth condition in both the simple environment and the complex environment. The contrast between the same-depth and different-depth conditions was significant across all combinations of task type and environment. With respect to physical environment complexity, the difference between simple and complex environments was non-significant only in the single-task different-depth condition. In all other combinations of task type and virtual depth, reaction times were significantly longer in the complex environment than in the simple environment. The details of the pairwise comparisons between repeated spatial configurations are provided in Table~\ref{tab:RT_Emmeans_Repeated} of Appendix~\ref{appendix:pairwise_comparison}.

Figure~\ref{fig:Reaction times for spatial configurations} illustrates the mean reaction times with their standard errors across task type, physical environment complexity, virtual element depth, and spatial configuration. For novel spatial configurations, the highest mean reaction time was observed in the complex environment under dual-task conditions with different depth, whereas the lowest mean reaction time occurred in the simple environment under single-task conditions with same depth. For repeated spatial configurations, the highest mean reaction time was found in the complex environment under dual-task conditions with different depth, while the lowest mean reaction time was observed in the simple environment under single-task conditions with same depth.

\begin{figure*}[tbp]
  \centering
  \includegraphics[width=0.97\textwidth]{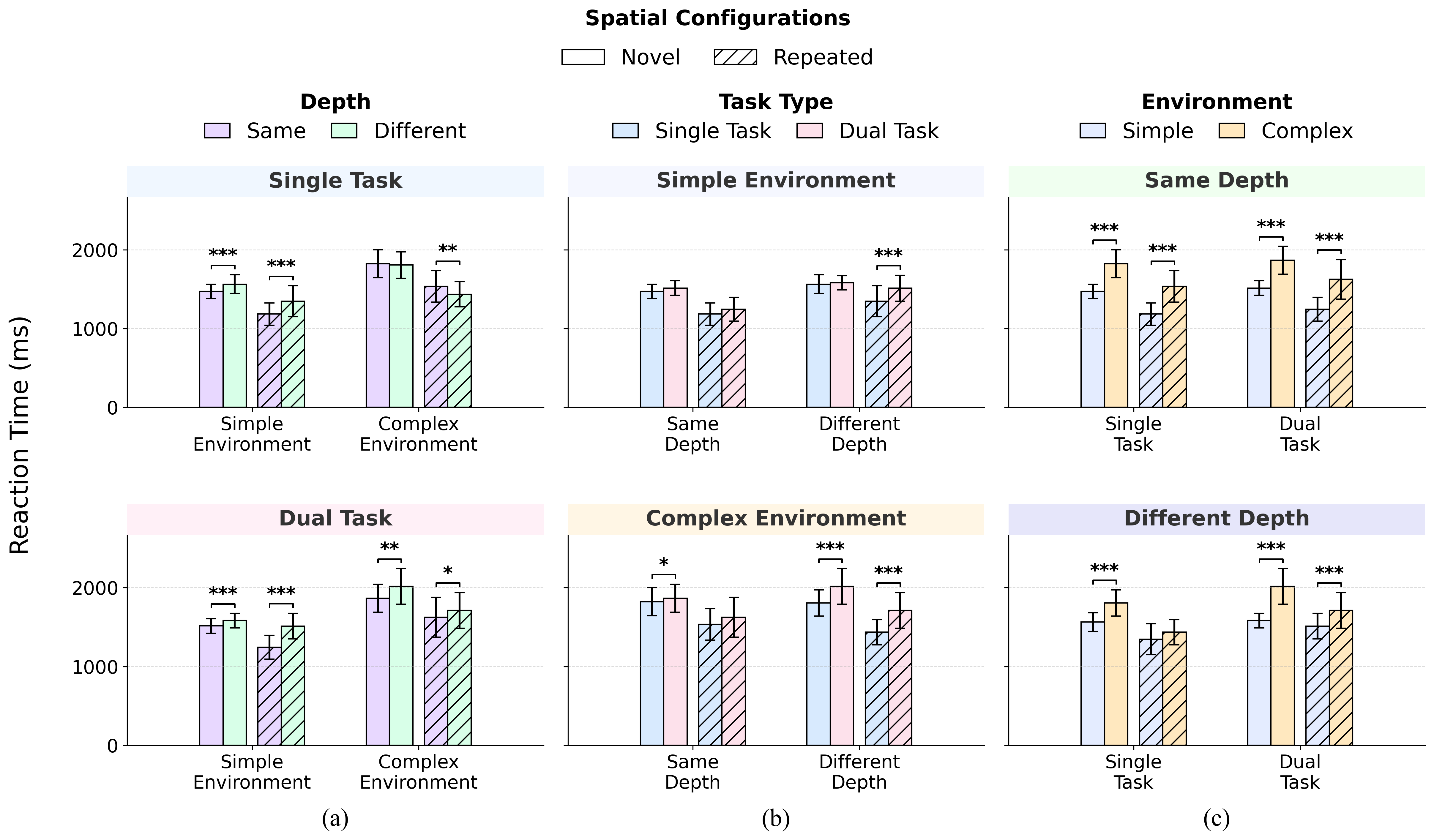}
  \caption{Bar plots for the mean reaction times (ms) of correct trials across three experimental factors: (a) task type, (b) virtual element depth, and (c) physical environment complexity, examined separately for novel and repeated spatial configurations. Statistical significance was assessed using pairwise comparisons of estimated marginal means with Tukey-adjusted p-values. Asterisks denote significant pairwise comparisons (* \textit{p} < .05, ** \textit{p} < .01, *** \textit{p} < .001).}
  \Description{This figure shows bar plots of the mean reaction times to our contextual cueing task. Reaction time was measured in milliseconds from the onset of the target to the participant's correct response, with incorrect trials excluded from the averages. The bar plots show the mean reaction times across three experimental factors: (a) task type, (b) virtual element depth, and (c) physical environment complexity, with results examined separately for novel and repeated spatial configurations. Statistical significance was assessed using pairwise comparisons of estimated marginal means with Tukey-adjusted p-values. Asterisks denote significant pairwise comparisons (* \textit{p} < .05, ** \textit{p} < .01, *** \textit{p} < .001).}
  \label{fig:Reaction times for spatial configurations}
\end{figure*}

\subsubsection{Generalized Linear Mixed-Effects Model for Keypress Accuracy}
We calculated the keypress accuracy as the proportion of incorrect key presses relative to the total number of keypresses. We analyzed it using a binomial GLMM with a logit link, where each trial was coded as either a correct response or an incorrect response. The fixed effects consisted of physical environment complexity, virtual element depth, task type, and spatial configurations, together with their interactions. We included random intercepts for participants to account for stable differences in error tendency across individuals. The reference condition was defined as the simple environment, same-depth virtual elements, a single-task setting, and novel spatial configurations, such that the intercept reflects the log-odds of correct responses under this condition.

Table~\ref{tab:Error Rate GLMM} reports the effects of the experimental factors from the GLMM on keypress on keypress accuracy. Positive unstandardized regression coefficient ($B$) indicate higher accuracy, whereas negative $B$ correspond to lower accuracy relative to the reference condition. Odds ratios (OR) denotes the effect size estimated by the GLMM~\cite{ben2020effectsize}. The variance of the random intercept was 1.181, with a standard deviation of 1.087. Compared to the simple environment, participants committed significantly more errors in the complex environment ($B = -0.660$, OR = 0.716, SE = 0.186, z = -3.559, \textit{p} = 3.73E$-4$). Main effects of virtual element depth, task type and spatial configurations were not statistically significant.

In terms of interactions, repeated spatial configurations significantly mitigated the detrimental effect of complex environments ($B = 1.115$, OR = 1.289, $SE = 0.464$, z = 2.402, \textit{p} = 1.63E$-2$). At higher orders, neither the three-way nor the four-way interactions yielded significant effects.

\begin{table*}[tbp]
\centering
\caption{Results of the generalized linear mixed-effects model predicting keypress accuracy from environmental complexity, virtual element depth, task type, and configuration type, with random intercepts for participants. The results showed that accuracy was significantly reduced in the complex environment. The only significant interaction emerged between the dual-task condition and repeated spatial configurations.}
\label{tab:Error Rate GLMM}
\begin{tabular*}{\textwidth}{@{\extracolsep{\fill}}lccccc@{}}
\toprule
\textbf{Parameter} & \textbf{$B$} & \textbf{OR} & \textbf{Std. Error} & \textbf{z value} & \textbf{p-value} \\
\midrule
(Intercept) & 4.509 & 61.577 & 0.272 & 16.556 & 1.44E-61*** \\
Complex Environment & -0.660 & 0.716 & 0.186 & -3.559 & 3.73E-04*** \\
Different Depth & 0.219 & 1.027 & 0.224 & 0.979 & 3.28E-01 \\
Dual Task & -0.299 & 0.898 & 0.199 & -1.502 & 1.33E-01 \\
Repeated Configuration & -0.483 & 0.975 & 0.281 & -1.721 & 8.53E-02 \\
Complex Environment × Different Depth & -0.229 & 0.931 & 0.272 & -0.844 & 3.99E-01 \\
Complex Environment × Dual Task & 0.267 & 1.055 & 0.251 & 1.067 & 2.86E-01 \\
Different Depth × Dual Task & -0.047 & 0.975 & 0.295 & -0.158 & 8.75E-01 \\
Complex Environment × Repeated Configuration & 0.088 & 1.013 & 0.349 & 0.251 & 8.02E-01 \\
Different Depth × Repeated Configuration & 0.190 & 1.035 & 0.432 & 0.440 & 6.60E-01 \\
Dual Task × Repeated Configuration & 1.115 & 1.289 & 0.464 & 2.402 & 1.63E-02* \\
Complex Environment × Different Depth × Dual Task & -0.109 & 0.986 & 0.362 & -0.302 & 7.63E-01 \\
Complex Environment × Different Depth × Repeated Configuration & 0.256 & 1.157 & 0.537 & 0.477 & 6.33E-01 \\
Complex Environment × Dual Task × Repeated Configuration & -0.704 & 0.910 & 0.560 & -1.256 & 2.09E-01 \\
Different Depth × Dual Task × Repeated Configuration & -0.832 & 0.882 & 0.643 & -1.293 & 1.96E-01 \\
Complex Environment × Different Depth × Dual Task × Repeated Configuration & 0.651 & 1.085 & 0.791 & 0.823 & 4.10E-01 \\
\bottomrule
\end{tabular*}
\begin{center}
\footnotesize
\textbf{Note.} $B$ denotes unstandardized regression coefficients, and OR denotes odds ratios.

Significance: * \textit{p} < .05, ** \textit{p} < .01, *** \textit{p} < .001.
\end{center}
\end{table*}

\begin{figure*}[tbp]
  \centering
  \includegraphics[width=0.97\textwidth]{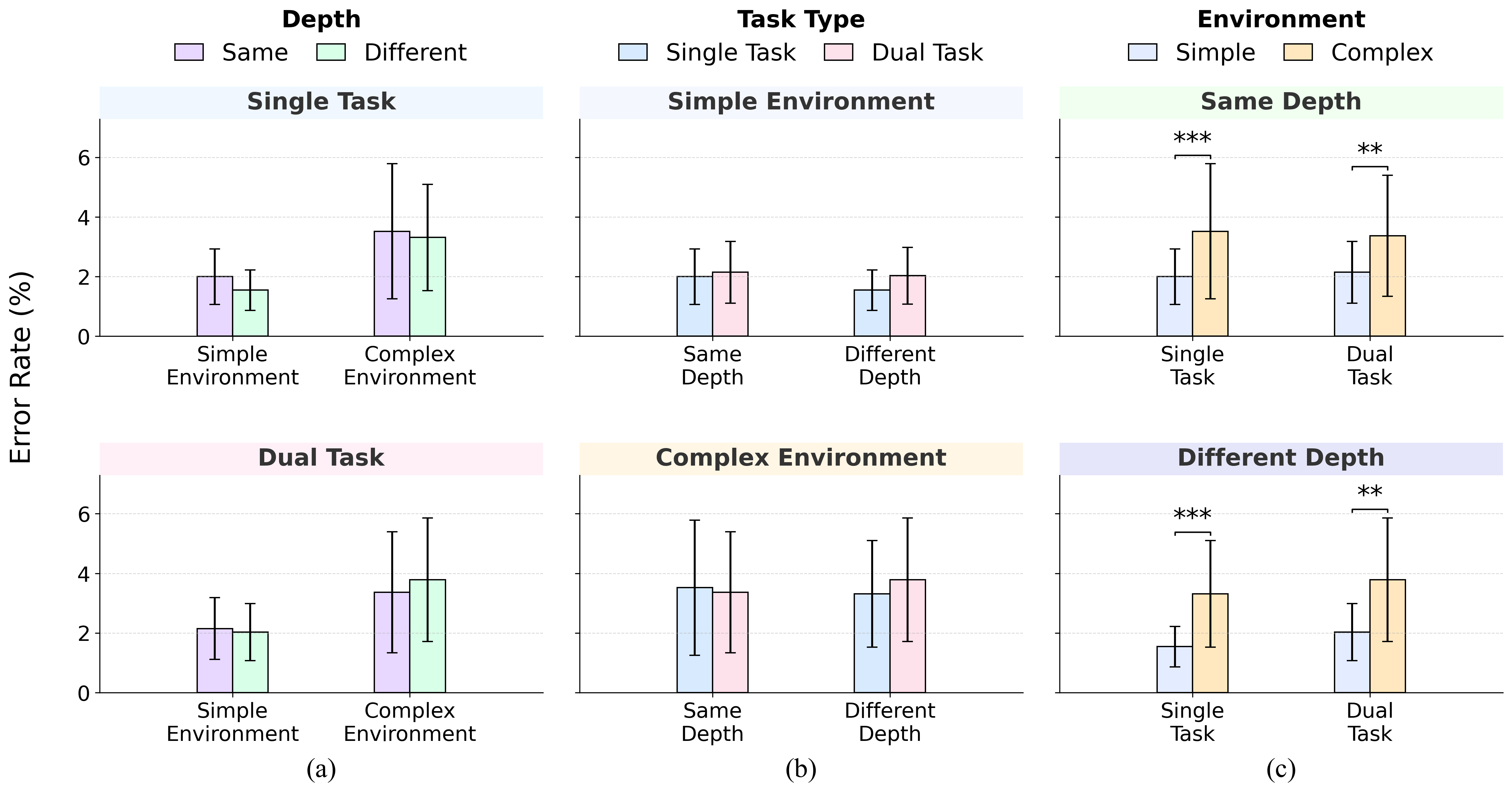}
  \caption{Bar plots for the mean error rates (\%) for keypress accuracy across three experimental factors: (a) task type, (b) virtual element depth, and (c) physical environment complexity. Statistical significance was assessed using pairwise comparisons of estimated marginal means with Tukey-adjusted p-values. Asterisks denote significant pairwise comparisons (* \textit{p} < .05, ** \textit{p} < .01, *** \textit{p} < .001). In both the different-depth and same-depth conditions, error rates were significantly higher in complex than in simple environments for both single-task and dual-task conditions. No other significant comparisons were found.}
  \Description{This figure shows bar plots of the mean error rates to our contextual cueing task. Error rates were measured by the percentage of incorrect responses to the total number of responses, reflecting keypress accuracy across conditions. The bar plots show the mean error rates across three experimental factors: (a) task type, (b) virtual element depth, and (c) physical environment complexity. Statistical significance was assessed using pairwise comparisons of estimated marginal means with Tukey-adjusted p-values. Asterisks denote significant pairwise comparisons (* \textit{p} < .05, ** \textit{p} < .01, *** \textit{p} < .001). In both the different-depth and same-depth conditions, error rates were significantly higher in complex than in simple environments for both single-task and dual-task conditions. No other significant comparisons were found.}
  \label{fig:Error Rate}
\end{figure*}

\subsubsection{Pairwise Comparisons of Keypress Accuracy}
We first compared the keypress accuracy, based on model-based EMMs from GLMM computed across combined spatial configurations. Dual-task performance did not differ significantly from single-task performance in any condition. Moreover, keypress accuracy in the different-depth condition did not differ significantly from that in the same-depth condition in any condition. However, keypress accuracy in the complex environment was significantly lower than in the simple environment across all combinations of task type and virtual depth. The details of the pairwise comparisons between combined spatial configurations are provided in Table~\ref{tab:Accuracy Emmeans All} of Appendix~\ref{appendix:pairwise_comparison}.

Figure~\ref{fig:Error Rate} illustrates the mean keypress error rates with their standard errors across the eight experimental conditions. The mean error rates were higher in the complex than in the simple environment, while the effects of virtual element depth and task type were comparatively modest. The lowest mean error rate was recorded in the simple environment under single task with different depth. The highest mean error rate in the complex environment under dual task with different depth.

Then, we compared keypress accuracy for novel spatial configurations. The comparison between the dual-task and single-task conditions was not significant in any combination of virtual element depth, and physical environment complexity. The contrast between the different-depth condition and the same-depth condition was also not significant in any combination. With respect to physical environment complexity, keypress accuracy in the complex environment was significantly lower than in the simple environment in all combinations of task type and virtual element depth. The details of the pairwise comparisons between novel spatial configurations are provided in Table~\ref{tab:Accuracy_Emmeans_Novel} of Appendix~\ref{appendix:pairwise_comparison}.

Finally, we compared keypress accuracy for repeated spatial configurations. The comparison between the dual-task and single-task conditions was not significant in any combination of virtual element depth, and physical environment complexity. The contrast between the different-depth condition and the same-depth condition was also not significant in any combination. Regarding physical environment complexity, keypress accuracy in the complex environment was significantly lower than in the simple environment only in the same-depth dual-task condition, and was not significantly different between environments in the other conditions. The details of the pairwise comparisons between novel spatial configurations are provided in Table~\ref{tab:Accuracy_Emmeans_Repeated} of Appendix~\ref{appendix:pairwise_comparison}.

Figure~\ref{fig:Error Rate Spatial Configurations} illustrates the mean keypress error rates with their standard errors across task type, physical environment complexity, virtual element depth, and spatial configuration. For novel spatial configurations, the highest mean error rate was observed in the complex environment under dual-task conditions with different depth, whereas the lowest mean error rate occurred in the simple environment under single-task conditions with different depth. For repeated spatial configurations, the highest mean error rate was in the complex environment under single-task conditions with different depth, while the lowest mean error rate was observed in the simple environment under dual-task conditions with same depth.

\begin{figure*}[tbp]
  \centering
  \includegraphics[width=0.97\textwidth]{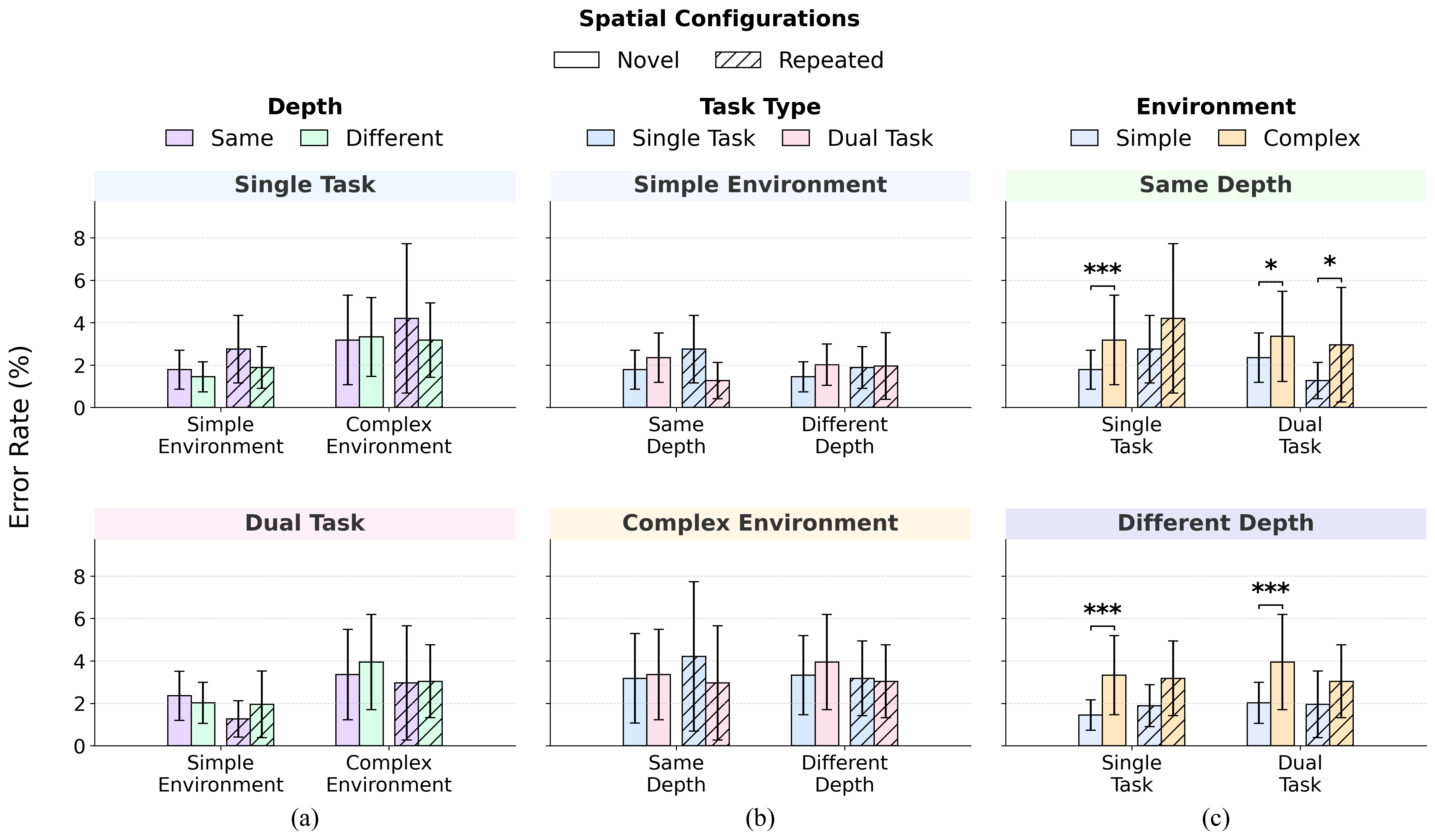}
  \caption{Bar plots for the mean error rates (\%) across three experimental factors: (a) task type, (b) virtual element depth, and (c) physical environment complexity, examined separately for novel and repeated spatial configurations. Statistical significance was assessed using pairwise comparisons of estimated marginal means with Tukey-adjusted p-values. Asterisks denote significant pairwise comparisons (* \textit{p} < .05, ** \textit{p} < .01, *** \textit{p} < .001).}
  \Description{This figure shows bar plots of the mean error rates to our contextual cueing task. Error rates were measured by the percentage of incorrect responses to the total number of responses, reflecting keypress accuracy across conditions. The bar plots show the mean error rates of keypresses across three experimental factors: (a) task type, (b) virtual element depth, and (c) physical environment complexity, with results separated for novel and repeated spatial configurations. Statistical significance was assessed using pairwise comparisons of estimated marginal means with Tukey-adjusted p-values. Asterisks denote significant pairwise comparisons (* \textit{p} < .05, ** \textit{p} < .01, *** \textit{p} < .001).}
  \label{fig:Error Rate Spatial Configurations}
\end{figure*}

\subsection{Spatial Regularity Memory}
\label{Result: Spatial Regularity}
For the contextual cueing task, we conducted pairwise comparisons based on the model-based EMMs from LMM to examine whether reaction times in the novel spatial configurations were greater than those in the repeated spatial configurations. A significant result would indicate that repeated spatial configurations facilitated faster visual search, indicating memorisation of spatial regularities. Across all conditions, participants responded significantly faster to repeated spatial configurations as compared to novel spatial configurations, suggesting that they exploited spatial regularities to improve search efficiency. The results of pairwise comparisons are detailed in Table~\ref{tab:RT_Emmeans_SpatialConfig}. Additional plots are provided in Figure~\ref{fig:One-Tailed} of Appendix~\ref{appendix:one-tailed}.

\begin{table*}[tbp]
\centering
\caption{Results of pairwise model-based estimated marginal means from LMM comparing novel versus repeated spatial configurations, reported separately for physical environment complexity, virtual depth, and task type. Positive estimates indicate longer reaction times for novel relative to repeated spatial configurations.}
\label{tab:RT_Emmeans_SpatialConfig}
\begin{tabular*}{\textwidth}{@{\extracolsep{\fill}}ccccccc@{}}
\toprule
\multicolumn{3}{c}{\textbf{Condition}} & \multicolumn{4}{c}{\textbf{Result}} \\
\cmidrule(lr){1-3} \cmidrule(lr){4-7}
\textbf{Environment} & \textbf{Depth} & \textbf{Task Type} & \textbf{Estimate} & \textbf{$t(23349)$} & \textbf{$P_{\text{adjusted}}$} & \textbf{Cohen's $d$} \\
\midrule
Simple Environment  & Same Depth    & Single Task &  0.218 &   9.824 & 4.89E-23*** &  0.454 \\
Simple Environment   & Same Depth      & Dual Task   &  0.190 &   8.575 & 5.25E-18*** &  0.394 \\
Simple Environment   & Different Depth & Single Task &  0.155 &   7.005 & 1.27E-12*** &  0.323 \\
Simple Environment   & Different Depth & Dual Task   &  0.065 &   2.930 & 1.70E-03**  &  0.135 \\
Complex Environment & Same Depth    & Single Task &  0.156 &   6.925 & 2.24E-12*** &  0.323 \\

Complex Environment  & Same Depth     & Dual Task   &  0.138 &   6.194 & 2.98E-10*** &  0.286 \\

Complex Environment  & Different Depth & Single Task &  0.230 &  10.317 & 3.33E-25*** &  0.479 \\
Complex Environment & Different Depth & Dual Task   &  0.117 &   5.262 & 7.19E-08*** &  0.244 \\
\bottomrule
\end{tabular*}
\begin{center}
\footnotesize Significance: * \textit{p} < .05, ** \textit{p} < .01, *** \textit{p} < .001.
\end{center}
\end{table*}

\begin{figure*}[tbp]
    \centering
    \includegraphics[width=0.8\textwidth]{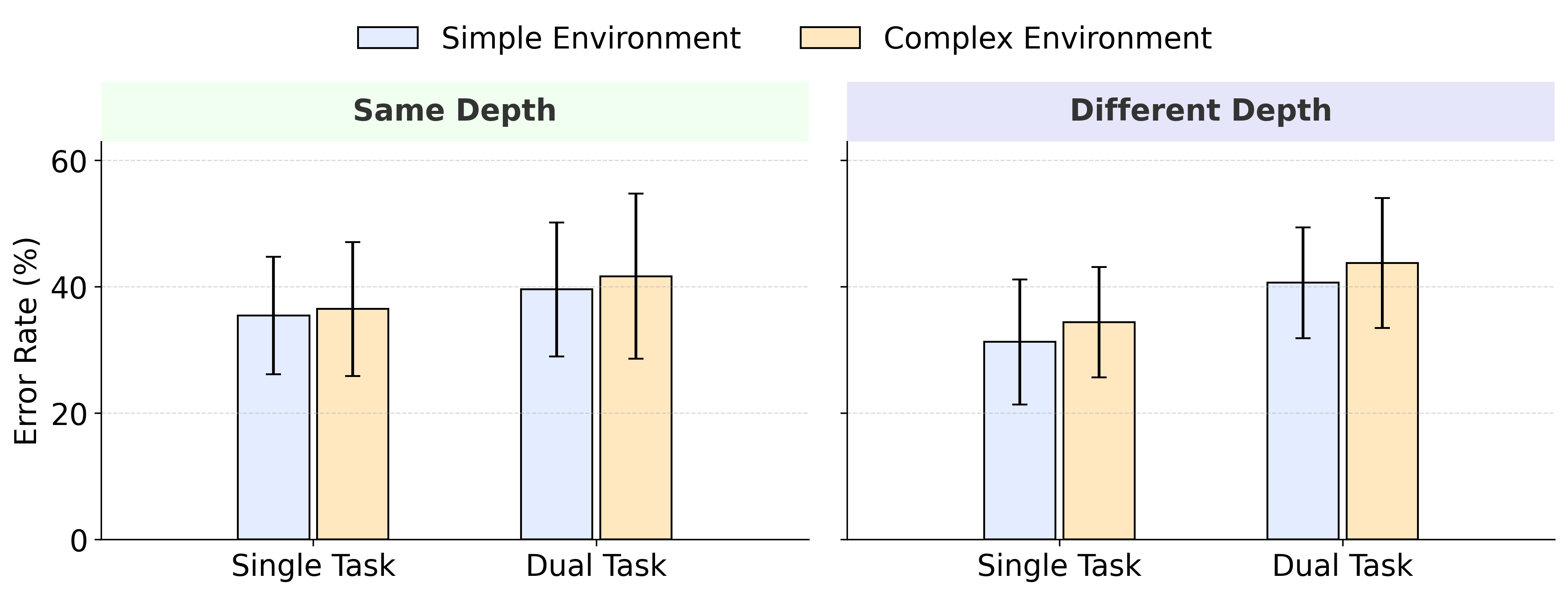}
    \caption{Bar plots of the mean recognition error rates (\%) with standard errors. The left panel shows results for the same-depth condition, and the right panel shows results for the different-depth condition. Within each panel, error rates are separated by physical environment complexity and task type. Higher error rates indicate greater difficulty in distinguishing repeated from novel spatial configurations.}
    \Description{This figure shows bar plots of the mean error rates to our recognition task. In the recognition task, participants were asked to judge whether a given spatial configuration of items had been repeated during the contextual cueing task or was newly generated. Error rates were measured by the percentage of incorrect responses, with higher error rates indicating greater difficulty in discriminating repeated from novel displays. The left panel shows results for the same-depth condition, and the right panel shows results for the different-depth condition. Within each panel, error rates are broken down by physical environment complexity (simple vs. complex) and task type (single-task vs. dual-task). Bars represent mean error rates.}
    \label{fig:Error Rate Recognition Task}
\end{figure*}

To examine whether participants’ memory for spatial regularities was explicit or implicit, we administered a recognition task. Figure~\ref{fig:Error Rate Recognition Task} illustrates participants’ mean error rates in the recognition task when distinguishing repeated from novel spatial configurations. Overall, the mean error rates ranged between 30\% and 45\%, indicating relatively low accuracy in this task. The lowest mean error rate was observed in the simple environment under single task with different depth, whereas the highest mean error rate was recorded in the complex environment under dual task with different depth.

\subsection{Perceived Workload}
We used a CLMM to investigate responses for each sub-scale in the NASA-TLX questionnaire. The fixed effects of the CLMM consisted of physical environment complexity, virtual element depth and task type, together with their interactions. Random intercepts were included for participants to to capture variability across individuals. We also reported Nagelkerke's pseudo-$R^2$ as an index of overall effect size for the CLMM. It is a likelihood-based index that rescales Cox--Snell's $R^2$ to the $0$--$1$ range and reflects the improvement in fit of the full model over a random-intercept null model~\cite{nagelkerke1991note}. The results indicates that under the dual-task condition, participants reported significantly higher ratings of mental demand (Estimate = 2.601, SE = 0.546, Nagelkerke's pseudo-$R^2$ = 0.400, \textit{p} = 1.88E$-6$), physical demand (Estimate = 2.242, SE = 0.6116, Nagelkerke's pseudo-$R^2$ = 0.248, \textit{p} = 2.46E$-4$), temporal demand (Estimate = 2.466, SE = 0.570, Nagelkerke's pseudo-$R^2$ = 0.301, \textit{p} = 1.5E$-5$), effort (Estimate = 2.979, SE = 0.562, Nagelkerke's pseudo-$R^2$ = 0.437, \textit{p} = 1.15E$-7$), frustration (Estimate = 3.238, SE = 0.662, Nagelkerke's pseudo-$R^2$ = 0.337, \textit{p} = 1.00E$-6$), and performance (Estimate = 1.781, SE = 0.569, Nagelkerke's pseudo-$R^2$ = 0.176, \textit{p} = 1.75E$-3$). The detailed participant-level random effects for the CLMMs are reported in Table~\ref{tab:clmm_random_effects} of Appendix~\ref{appendix:perceived_workload}. In contrast, neither physical environmental complexity nor virtual element depth, nor the interactions showed significant effects on any of the individual NASA-TLX sub-scales. The detailed mean perceived workload scores are reported in Table~\ref{tab:nasa_tlx_by_condition_compact} of Appendix~\ref{appendix:perceived_workload}.

To better present the significant effects of the experimental factors on the dependent variables, Table~\ref{tab:main_effects} summarises the significant main effects on reaction times, keypress accuracy, and perceived workload, and Table~\ref{tab:interaction_effects} summarises the significant interaction effects on these measures.

\begin{table*}[tbp]
  \centering
  \caption{Summary of significant main effects on reaction times, keypress accuracy, and perceived workload.}
  \label{tab:main_effects}

  \begin{tabular*}{\textwidth}{@{\extracolsep{\fill}}p{0.5\textwidth}ccc@{}}
    \toprule
    \textbf{Condition} & \textbf{Reaction Times} & \textbf{Keypress Accuracy} & \textbf{Perceived Workload} \\
    \midrule
    Complex environment    & \up          & \down             & -- \\
    Different depth        & \up          & \down             & -- \\
    Dual task              & --           & --                & \up \\
    Repeated configuration & \down        & --                & -- \\
    \bottomrule
  \end{tabular*}

  \begin{minipage}{\textwidth}
    \footnotesize \textbf{Note.} An orange upward arrow (\up{}) indicates an increase relative to the baseline condition, a blue downward arrow (\down{}) indicates a decrease relative to the baseline condition, and -- indicates that no substantial effect was observed.
  \end{minipage}

\end{table*}

\begin{table*}[tbp]
  \centering
  \caption{Summary of significant interaction effects on reaction times, keypress accuracy, and perceived workload.}
  \label{tab:interaction_effects}

  \begin{tabular*}{\textwidth}{@{\extracolsep{\fill}}p{0.5\textwidth}ccc@{}}
    \toprule
    \textbf{Condition} & \textbf{Reaction Times} & \textbf{Keypress Accuracy} & \textbf{Perceived Workload} \\
    \midrule
    Complex Environment $\times$ Different Depth                                   & \down & --  & -- \\
    Complex Environment $\times$ Repeated Configuration                            & \up   & --  & -- \\
    Different Depth $\times$ Repeated Configuration                                & \up   & --  & -- \\
    Dual Task $\times$ Repeated Configuration                                      & --    & \up & -- \\
    Complex Environment $\times$ Different Depth $\times$ Repeated Configuration   & \down & --  & -- \\
    \bottomrule
  \end{tabular*}

  \begin{minipage}{\textwidth}
    \footnotesize \textbf{Note.} Arrows indicate the direction of the estimated interaction coefficient in the mixed-effects models. An orange upward arrow (\up{}) denotes a positive interaction estimate (i.e., a larger outcome than expected from the additive main effects), a blue downward arrow (\down{}) denotes a negative interaction estimate, and -- denotes no statistically significant interaction.
  \end{minipage}
\end{table*}

\subsection{Qualitative Result}
The qualitative insights obtained from our post-experiment interview provides supplementary perspectives to our quantitative findings. We examined potential effects of the equipment, asking whether participants encountered any discomfort or difficulties when using the MR headset or keyboard. All participants confirmed that they did not experience such difficulties.

We then asked participants to identify the conditions under which it was easiest or most difficult to conduct visual search. Twenty-one participants (87.5\%) out of twenty-four participants reported that conducting visual search was easiest in the simple environment under single-task condition with virtual elements being presented at same depth, whereas only three participants (12.5\%) considered the simple environment with single task and different depth to be the most difficult. Seventeen participants (70.83\%) perceived the combination of complex environment, dual task, and different depth as the most challenging condition for visual search. 

We also asked our participants to elaborate on the sources of interference during the visual search task. Fourteen participants (58.33\%) highlighted environmental complexity, with one mentioning that the clutter in the complex environment substantially increased the difficulty of locating the virtual target ``T''. Nine participants (37.5\%) emphasized task type as the primary source of interference, with four participants (16.67\%) noting that it was ``impossible'' to divide attention between two tasks, as the secondary auditory task disrupted the visual search process and sometimes even caused them to ``forget'' about searching for the virtual target ``T''. Only one participant indicated that the placement of virtual objects at different depth levels was the primary factor affecting on the visual search performance. It is worth noting that all participants consistently acknowledged that the dual-task condition substantially hindered the visual search performance.

We further examined how different factors influenced memory for spatial regularities. Thirteen participants (54.17\%) reported that spatial configurations were easiest to remember in the simple environment with single task and same depth, seven participants (29.17\%) selected the simple environment with single task and different depth, and four participants (16.67\%) pointed to the complex environment under single-task conditions. Regarding the most difficult scenarios, thirteen participants (54.17\%) identified the complex environment with dual task and different depth, followed by six participants (25\%) who selected the complex environment with dual task and same depth.

When asked about the main factors influencing memory of spatial regularities, fifteen participants (62.5\%) identified task type, eight participants (33.33\%) pointed to environmental complexity, and only one participant emphasized the role of virtual object depth. Seven participants (29.17\%) noted that virtual elements were presented at different depth enhanced the sense of three-dimensionality in the simple environment, thereby facilitating memory. However, in the complex environment, the abundance of objects caused depth cues to hinder rather than facilitate the perceptibility of repeated spatial configurations. Opinions on environmental complexity were more divided: Two participants (8.3\%) reported that the arrangement of physical objects in the complex background provided cues that supported memory formation. Six participants (25\%) reported no noticeable differences in their ability to retain repeated spatial configurations between the complex and simple environments. In contrast, the remaining participants (66.7\%) indicated that the excessive amount of visual information made it difficult to retain repeated spatial configurations. Notably, all participants unanimously agreed that they were unable to explicitly recall the exact layout of repeated spatial configurations, though many reported a vague sense of familiarity with certain displays. Additional details of participants’ responses are provided in Appendix~\ref{appendix:interview_results}.

\section{Discussion}
The aim of this research is to understand how physical environment complexity, virtual element depth, and task type influences visual search and spatial regularity memory in MR. Overall, our findings demonstrate that these factors shape user performance and experience in distinct ways. Based on the LMM, GLMM, and CLMM results, the secondary auditory task did not have a significant main effect on visual search performance. Therefore, H1 was not supported, even though the secondary auditory task was perceived as the most demanding condition. In contrast, our results show that complex environments have the most adverse effects on visual search task,  consistent with H2. However, they were not perceived to significantly increase workload. Contrary to H3, different depths of virtual elements had a significant and negative main effect on visual search performance. In addition, virtual elements at different depths did not significantly increase perceived workload. These findings suggest a need to balance user perceptions with objective performance measures in visual MR tasks, such that MR solutions are optimized for both better user experience and user performance. Our pairwise comparison results also indicate that different combination of factors may exert unique effects on novel and repeated spatial configurations. This highlights the need to design around contextual factors when expecting MR use in familiar and unfamiliar layouts. Reaction times were significantly slower for repeated than for novel spatial configurations in all conditions, indicating that H4 was not supported. The low accuracy in the recognition task and participants’ reports in the post-experiment interviews suggest that spatial regularity memory is primarily implicit.

\subsection{Effect of Task Type, Physical Environment Complexity, and Virtual Element Depth on Visual Search}
The LMM and GLMM results indicated that the secondary auditory task did not have a significant main effect on visual search performance, diverging from prior reports of reduced visual performance in dual-task experiment~\cite{richard2002effect, gherri2011active}. Therefore, H1 is not supported. We speculate that this discrepancy may arise because prior studies focused on isolated factors and did not consider how multiple factors jointly shape dual-task costs. However, despite LMM results showing no significant global interaction effects between task type and the other independent variables, the follow-up pairwise comparisons indicate that task type influences visual search performance at specific combinations of physical environment complexity and virtual element depth. Specifically, although the secondary auditory task did not have a significant effect on visual search reaction times in simple environments with same-depth virtual elements, it had a statistically significant, but small-to-moderate effect, in visually complex environments under both the same-depth (Cohen’s d = 0.078, \textit{p} = 1.73E-02) and different-depth conditions (Cohen’s d = 0.287, \textit{p} = 2.55E-18), and in simple environments where virtual elements were presented at the different depth (Cohen’s d = 0.140, \textit{p} = 1.82E-05). This may be caused by differences in visual characteristics of virtual elements compared to physical objects---making the virtual target relatively easy to locate in visually simple environments, even when performing a secondary task. However, as complexity increases through physical clutter and the need to refocus across different virtual element depths, the search task becomes more difficult, and the addition of a secondary task further taxes limited cognitive resources~\cite{plass2010cognitive, sweller2011cognitive}. These results suggest that when MR applications are intended for use in complex real-world settings and require the distribution of virtual content across different depth planes, additional care is warranted to mitigate the heightened risk of dual-task interference. For instance, in MR simulations for excavation training~\cite{moore2019review, azzam2023mixed}, where virtual pipelines, exclusion zones, and grade lines are rendered across multiple depths, the system should defer non-urgent status notifications to prevent dual-task interference with visual search.

Moreover, our pairwise comparisons for novel spatial configurations indicate that the secondary auditory task significantly slowed reaction times only in complex environments under both the same-depth (Cohen’s d = 0.060, \textit{p} = 4.17E-02) and different-depth conditions (Cohen’s d = 0.169, \textit{p} = 8.47E-09), although the effect size was small in magnitude. This suggests that the higher perceptual and attentional load incurred by visually complex environments, combined with the uncertainty in visually searching unfamiliar configurations can exacerbate challenges introduced by a secondary task. In contrast, for repeated spatial configurations, the secondary auditory task significantly slowed reaction times only when virtual elements were presented at different depths, in both simple (Cohen’s d = 0.234, \textit{p} = 6.08E-05) and complex environments (Cohen’s d = 0.404, \textit{p} = 5.84E-12). This may be because part of the search is offloaded to memory for familiar layouts, so the limiting factor shifts from overall scene complexity to how easily these regularities can be retrieved and applied in three dimensions. When virtual elements are presented at different depths, exploiting repeated spatial configurations requires binding spatial relations across depth planes, which makes it harder for participants to use the learned spatial regularities. Consequently, the dual-task cost appears primarily in the different-depth condition. These findings imply that MR application designers need to consider not only how easily users can locate targets from the immediate visual scene, but also how effectively targets can be located by relying on memory for consistent spatial regularities.

Additionally, although the LMM and GLMM analyses revealed no significant main effect of the secondary auditory task on visual search performance, subjective reports were strikingly consistent but contrasting: interview responses indicated that all participants perceived the dual-task condition as substantially impairing visual search. NASA-TLX responses similarly exhibited increased workloads across all individual sub-scales for the dual-task condition when compared to single-task condition. This divergence highlights the need to balance users’ subjective experience with objective performance to ensure both usability and effectiveness. The result also highlight that assessments of MR applications should not rely exclusively on objective indicators such as reaction times or accuracy. Subjective measures of workload, comfort, and perceived usability are equally critical for capturing the full spectrum of user experience.

The LMM results also indicates that physical environment complexity had a significant effect on visual search performance, with complex environments markedly slowing participants’ reaction times ($\beta$ = 0.305, \textit{p} = 9.65E-29). While expected and aligned with H2, this result presents a key consideration for designing MR applications across different environmental contexts. For example, collaborative office spaces with no furniture and just whiteboards may exhibit less visual complexity, allowing for MR applications with more involved searching interactions. In contrast, navigation applications deployed in busy streets with pedestrians and cars require careful design that constrains both virtual element depth and task complexity.

An additional insight from our LMM analysis suggest that when virtual elements were presented at different depths, participants required more time to complete visual search, which is contrary to H3. The LMM results also indicates an interesting significant interaction effect, where varying virtual element depth can mitigate the negative impact of complex environments on visual search reaction times ($\beta$ = 0.110, \textit{p} = 4.54E-3). A possible explanation is that visual interference in complex environments stems from clutter that causes targets and distractors to be grouped together based on similarity and proximity~\cite{kohler1967gestalt, koffka2013principles}. When depth cues are introduced, they provide segmentation information, creating clearer boundaries between foreground and background and thereby reducing interference. However, In simple environments, there is relatively little clutter to segment, so additional depth variation does not offer substantial segmentation benefits. Instead, it requires participants to distribute attention across multiple depth planes, increasing the effort needed to locate the target and ultimately slowing visual search. This divergence suggests that virtual element depth functions as a double-edged sword. In simple environments, depth should be regarded both as a resource for conveying spatial structure and as a constraint when efficiency is critical. In complex environments, MR applications can leverage depth cues to reduce interference and support more efficient target detection.

The pairwise comparisons of reaction times to combined spatial configurations with different virtual element depths across all levels of physical environment complexity and task type provides further context. These post-hoc analysis reveal presenting virtual elements at different depths significantly reduced reaction times only in the complex environments under the single-task condition (Cohen’s d = -0.098, \textit{p} = 2.91E-03), albeit with a small effect size. In contrary, in the complex environment under the dual-task condition, virtual elements presented at different depths yielded significantly longer reaction times than virtual elements presented at the same depth (Cohen’s d = 0.110, \textit{p} = 7.47E-04). This may because the combination of a visually complex environment and a concurrent auditory task already consumes substantial cognitive resources, and processing virtual elements at different depths additionally requires integrating binocular disparity~\cite{qian1997binocular, read2006does, verhoef2016binocular}, perspective cues~\cite{harwerth1998effects, van1979interrelation}, and focal adjustments~\cite{lee2024effects, watt2005focus} to determine the target’s location, thereby intensifying competition for limited cognitive resources~\cite{kahneman1973attention, wickens2020processing} and slowing responses.

In visual search, speed is not the only critical measure, and accuracy is equally important. In many cases, rapid but erroneous selections, such as clicking on an incorrect icon and then correcting it, consume more time than slower but accurate target identification. This led us to question how contextual factors in MR influence visual search accuracy.

The GLMM results showed that complex environments significantly reduced the target search accuracy (OR = 0.716, \textit{p} = 3.73E-04), potentially due to clutter that obscured targets and distractors. In contrast, although presenting virtual elements at different depths slowed search speed, it did not significantly reduce accuracy. Similarly, adding a dual task had no significant effect on accuracy. These findings suggest that MR applications can tolerate depth variation and moderate multitasking without significantly compromising accuracy, but designers should be cautious in cluttered physical settings where visual confusion is most likely. For example, in a classrooms with multiple posters, charts, and student activity, virtual learning materials should be presented with simplified, highly distinctive cues to facilitate efficient information search. In contrast, in controlled study environments such as quiet self-study rooms with only desks and a few papers, MR application can place virtual element at different depths to enhance information density and add concurrent tasks (e.g., interactive questions) to enrich the learning experience.

\subsection{Effect of Task Type, Physical Environment Complexity, and Virtual Element Depth on Memory for Spatial Regularities}
Our LMM results indicate a significant main effect of spatial configurations on visual search performance. This is further supported by our pairwise comparisons between repeated and novel spatial configurations, which demonstrates that participants successfully exploited spatial regularities to accelerate search in all conditions. This finding is inconsistent with H4 and with prior studies suggesting that dual-task demands suppress the formation and expression of spatial regularity learning~\cite{cooper2025multitasking}. One possible explanation concerns the number and variety of repeated spatial configurations used in earlier work. Many of these studies employed a large set of distinct repeated spatial configurations, significantly increasing the complexity of learning spatial regularities. In such studies, participants may not have been given sufficient opportunity to register the different repeated spatial configuration. As such, the absence of the spatial regularity learning may be misattributed to a specific experimental factor rather than to insufficient opportunities for learning. These results highlight the need for future work to determine whether a failure to observe spatial regularity learning is due to an insufficient number of repetitions of repeated spatial configurations, or whether it reflects a genuine impact of the experimental conditions.

Additionally, our post-hoc pairwise comparisons reveal differences in the degree to which different spatial configurations influence visual search times across the different experimental conditions. Notably, minimal effect was observed in simple environments, where virtual targets were presented at different depths and a secondary task was present (Cohen’s d = 0.135, \textit{p} = 1.70E-03). We speculate that, in different-depth settings, spatial regularity learning requires encoding the depth relations between virtual elements. Complex environments provide abundant contextual cues that support such higher-order binding, whereas simple environments offer few contextual cues, making bindings fragile. The dual task further taxed working memory, straining capacity for encoding and consolidating these relational structures. This implies that MR applications can enhance spatial regularity learning by anchoring virtual elements to physical objects in the environment. For example, a virtual TV guide may be aligned with the edge of a physical television screen, and a speech-assistant notification could appear near the speaker. Such consistent spatial associations enable users to acquire and exploit spatial regularities more intuitively.

Although spatial regularities facilitated faster search, participants consistently reported being unable to recall repeated spatial configurations during the interview. Taken together with the high error rates in the recognition task, these findings suggest that the acquisition of spatial regularities relied primarily on implicit memory rather than explicit memory. These results highlight the importance of repetition and consistency in MR spatial layouts, allowing users to benefit from spatial regularities even when explicit memory is limited.

The LMM results also showed that complex environments significantly reduced the facilitation effect of repeated spatial configurations ($\beta$ = 0.122, \textit{p} = 4.71E-2). The presence of numerous, and potentially salient, physical objects draws attentional resources~\cite{desimone1995neural} away from virtual targets and makes it harder for participants to exploit repeated spatial configurations during search. In addition, users exhibited significantly slower search times to repeated spatial configurations with virtual elements at different depths ($\beta$ = 0.123, \textit{p} = 4.47E-2). When complex environments were combined with different-depth conditions, the facilitation effect of repeated spatial configurations was restored and even enhanced ($\beta$ = -0.268, \textit{p} = 2.00E-3). A likely explanation is that depth cues stratified virtual and physical elements, making target–distractor configurations emerge more distinctly from clutter and thereby supporting the use of spatial regularities. These results indicate that MR applications can enhance spatial regularity learning by aligning virtual content with the depth structure of the real environment, thereby minimizing interference from physical clutter and facilitating more efficient search. For instance, in a learning setting, virtual annotations or immediate prompts can be attached to notebooks or the desktop surface, while virtual search results or learning supplementary materials can be placed at farther depth levels near the monitor. Such spatial organization helps users distinguish between ``nearby task-relevant information'' and ``distant reference information''. By aligning virtual content with the depth structure of the real environment, users can rely on familiar spatial anchors to organize virtual information, which in turn simplifies visual search.

In addition, our GLMM results suggest that repeated spatial configurations reduced the negative impact of dual-task demands on accuracy (OR = 1.289, \textit{p} = 1.63E-02). Repeated configurations increase the robustness of MR interactions in dual task contexts, enabling users to maintain accuracy even when attention is divided. This indicates that when multitasking in MR applications is unavoidable, maintaining stable and easily learnable spatial regularities, such as preserving the same spatial layout of virtual elements across contexts, can mitigate performance costs imposed by multitasking demands.

\subsection{Implications Beyond MR}

As interactions with many digital user interfaces often involve visual search~\cite{putkonen2025understanding, bunian2021vins}, it is important to consider our findings in broader HCI contexts. Variable depth cues are common in 3D modelling applications and games~\cite{mehrabi2013making}, and the placement of interface elements can aid or hinder search based on users' prior experiences with similar layouts~\cite{todi2019individualising}. For example, a desktop 3D interior design study might evaluate whether distributing interface elements across depth improves visual search and editing efficiency. Different interface layouts may be tested, where modelling tools and visual aids are placed in varying depths---such as keeping key parameters in a near-camera panel while embedding context-sensitive controls and labels as floating widgets anchored to objects in the 3D scene. If the study repeatedly uses the same room template and highly similar editing tasks, users may appear to search more efficiently over time because they learn the recurring depth-structured placement of tools and cues. With repeated exposure, novices can learn not only where controls are located on the screen but also which depth plane they typically occupy, while experts may benefit earlier if these conventions match prior CAD workflows. In this case, improved visual search performance may reflect familiarity with a recurring depth-based layout rather than the causal effect of the layout manipulation itself.

Moreover, understanding how the placement of virtual elements can better support visual search is vital for many applications, as design choices about where information appears, can either guide attention efficiently or impose additional visual search costs~\cite{wolfe2021guided, wolfe2020visual}. The relationship between virtual element depth and physical environment complexity can also extend to other HCI contexts. For example, in in-vehicle head-up displays, designers may present virtual cues such as navigation guidance, hazard alerts, or speed-limit information within the driver’s forward view while the driver simultaneously processes a visually complex road scene. In demanding environments such as dense traffic, complex intersections, or poor visibility, the road scene can be visually cluttered, and critical cues may compete with background objects and motion for attention. In such contexts, distributing cues across depth planes can help visually separate them from the background and from one another, supporting faster target detection. For instance, designers could render the most urgent alerts (e.g., collision or hazard warnings) on a nearer depth plane, while placing lower-urgency guidance (e.g., navigation prompts) on a slightly farther plane, so that critical cues stand out without visually crowding the forward view.

Furthermore, our study suggests that HCI evaluations should consider not only objective performance measures but also users’ subjective experience, as subjective costs can reveal trade-offs that are not captured by performance metrics alone.

\section{Limitations and Future Work}
We acknowledge that our study has several limitations. First, the physical environment in our study was controlled and remained constant, whereas MR applications in real-world use often occur in dynamic and variable contexts~\cite{li2025noise}. For example, users may interact with MR while walking or when switching between home and office settings. Such environmental variability may undermine the stability of regularity cues and alert the visual search process~\cite{zellin2013here, zellin2014long}. Future work needs to examine how spatial regularities are learned and exploited under more naturalistic and dynamic environments.

Second, our study provides insights on the effects of MR factors specific to physical environment complexity, virtual element depth, dual-task presence, and spatial regularities on visual search performance. However, other potentially relevant factors were standardized, such as virtual element color and size, to minimize potential confounding factors. In real-world MR applications, virtual elements often vary in color, size, and shape. For instance, navigation arrows may be much larger than textual prompts, and educational applications may include vividly colored models that strongly contrast with their surroundings. These factors may influence visual search efficiency and spatial regularity memory. However, examining all relevant factors was beyond the scope of this paper and future research is needed to examine the potential influence of additional factors on visual search efficiency and spatial regularity memory in MR.

Third, we based our sample size on prior visual search ~\cite{makovski2016context, chen2024representation} and related HCI studies~\cite{caine2016local}. However, we acknowledge that an a priori power analysis would have provided us with a more rigorous basis for determining the required sample size.

\section{Conclusion}
In this work, we investigated the effects of physical environment complexity, virtual element depth, and dual-task presence on visual search performance and spatial regularity memory in MR. In contrast to prior work, our findings indicate that the interference from a secondary tone-counting task on visual search in MR was dependent on environmental and virtual element depth. Furthermore, differences between participants’ objective and subjective assessments underscore the need to balance efficiency with user experience. The differing factor effects for novel and repeated spatial configurations also highlight the need to address both efficient target localisation in the immediate visual scene and the effective exploitation of memory for consistent spatial regularities in MR interface design. Although memory for spatial regularities is implicit, it can significantly facilitate visual search. Our results also highlight the need to distinguish absences of spatial regularity learning due to limited repetitions from those caused by the experimental manipulations. Our findings contribute towards better understanding of visual search and spatial regularity memory in MR by disentangling the respective influences of physical environment complexity, virtual element depth and task demands.

\begin{acks}
We sincerely thank our participants for their time and the reviewers for their feedback, which helped improve this paper. We also thank the members of the AID-Lab for their assistance in various ways.
\end{acks}

\bibliographystyle{ACM-Reference-Format}
\bibliography{sample-base}

\clearpage

\appendix
\onecolumn
\section{Supplementary Tables of Pairwise Comparisons}
\label{appendix:pairwise_comparison}

\begin{table*}[hbp]
\centering
\footnotesize
\caption{Pairwise comparisons of estimated marginal means of log reaction times across combined spatial configurations, separately for task type, virtual depth, and physical environmental complexity. Positive estimates indicate longer reaction times in Level~1 relative to Level~2.}
\label{tab:RT Emmeans All}
\begin{tabular*}{\textwidth}{@{\extracolsep{\fill}}ccccccccc@{}}
\toprule
\multicolumn{2}{c}{\textbf{Comparison}} & \multicolumn{3}{c}{\textbf{Condition}} & \multicolumn{4}{c}{\textbf{Result}} \\
\cmidrule(lr){1-2} \cmidrule(lr){3-5} \cmidrule(lr){6-9}
\textbf{Level 1} & \textbf{Level 2} & \textbf{Environment} & \textbf{Depth} & \textbf{Task Type} & \textbf{Estimate} & \textbf{$t(23349)$} & \textbf{$P_{\text{adjusted}}$} & \textbf{Cohen's $d$} \\
\midrule
\multicolumn{9}{l}{\textit{Task type comparisons}} \\
Dual Task   & Single Task & Simple Environment  & Same Depth     & --          &  0.027 &  1.748 & 8.05E-02   &  0.057 \\
Dual Task   & Single Task & Simple Environment & Different Depth & --          &  0.067 &  4.287 & 1.82E-05***&  0.140 \\
Dual Task   & Single Task & Complex Environment & Same Depth     & --          &  0.038 &  2.381 & 1.73E-02*  &  0.078 \\
Dual Task   & Single Task & Complex Environment & Different Depth & --          &  0.138 &  8.737 & 2.55E-18***&  0.287 \\
\addlinespace
\multicolumn{9}{l}{\textit{Depth comparisons}} \\
Different Depth  & Same Depth & Simple Environment  & -- & Single Task &  0.078 &  4.981 & 6.38E-07*** &  0.162 \\
Different Depth  & Same Depth & Simple Environment  & -- & Dual Task   &  0.118 &  7.536 & 5.03E-14*** &  0.245 \\
Different Depth  & Same Depth & Complex Environment & -- & Single Task & -0.047 & -2.977 & 2.91E-03**  & -0.098 \\
Different Depth  & Same Depth & Complex Environment & -- & Dual Task   &  0.053 &  3.372 & 7.47E-04*** &  0.110 \\
\addlinespace
\multicolumn{9}{l}{\textit{Environment comparisons}} \\
Complex Environment & Simple Environment & -- & Same Depth      & Single Task &  0.188 &  11.890 & 1.65E-32*** &  0.391 \\
Complex Environment & Simple Environment & -- & Same Depth      & Dual Task   &  0.198 &  12.629 & 1.93E-36*** &  0.412 \\
Complex Environment & Simple Environment & -- & Different Depth & Single Task &  0.063 &   3.975 & 7.05E-05*** &  0.130 \\
Complex Environment & Simple Environment & -- & Different Depth & Dual Task   &  0.133 &   8.483 & 2.33E-17*** &  0.277 \\
\bottomrule
\end{tabular*}
\begin{center}
\footnotesize Significance: * \textit{p} < .05, ** \textit{p} < .01, *** \textit{p} < .001.
\end{center}
\end{table*}

\begin{table*}[hbp]
\centering
\footnotesize
\caption{Pairwise comparisons of estimated marginal means of log reaction times across novel spatial configurations, separately for task type, virtual depth, and physical environmental complexity. Positive estimates indicate longer reaction times in Level~1 relative to Level~2.}
\label{tab:RT_Emmeans_Novel}
\begin{tabular*}{\textwidth}{@{\extracolsep{\fill}}ccccccccc@{}}
\toprule
\multicolumn{2}{c}{\textbf{Comparison}} & \multicolumn{3}{c}{\textbf{Condition}} & \multicolumn{4}{c}{\textbf{Result}} \\
\cmidrule(lr){1-2} \cmidrule(lr){3-5} \cmidrule(lr){6-9}
\textbf{Level 1} & \textbf{Level 2} & \textbf{Environment} & \textbf{Depth} & \textbf{Task Type} & \textbf{Estimate} & \textbf{$t(23349)$} & \textbf{$P_{\text{adjusted}}$} & \textbf{Cohen's $d$} \\
\midrule
\multicolumn{9}{l}{\textit{Task type comparisons}} \\
Dual Task   & Single Task & Simple Environment  & Same Depth      & --          &  0.013 &  0.930 & 3.50E-01        &  0.027 \\
Dual Task   & Single Task & Simple Environment  & Different Depth & --          &  0.022 &  1.570 & 1.18E-01        &  0.046 \\
Dual Task   & Single Task & Complex Environment & Same Depth      & --          &  0.029 &  2.040 & 4.17E-02*       &  0.060 \\
Dual Task   & Single Task & Complex Environment & Different Depth & --          &  0.081 &  5.760 & 8.47E-09***     &  0.169 \\
\addlinespace
\multicolumn{9}{l}{\textit{Depth comparisons}} \\
Different Depth  & Same Depth & Simple Environment  & -- & Single Task &  0.047 &  3.333 & 8.61E-04***     &  0.097 \\
Different Depth  & Same Depth & Simple Environment  & -- & Dual Task   &  0.055 &  3.955 & 7.69E-05***     &  0.115 \\
Different Depth  & Same Depth & Complex Environment & -- & Single Task & -0.010 & -0.691 & 4.90E-01        & -0.020 \\
Different Depth  & Same Depth & Complex Environment & -- & Dual Task   &  0.043 &  3.036 & 2.40E-03**      &  0.089 \\
\addlinespace
\multicolumn{9}{l}{\textit{Environment comparisons}} \\
Complex Environment & Simple Environment & -- & Same Depth      & Single Task &  0.157 &  11.138 & 9.65E-29***     &  0.325 \\
Complex Environment & Simple Environment & -- & Same Depth      & Dual Task   &  0.172 &  12.221 & 3.04E-34***     &  0.358 \\
Complex Environment & Simple Environment & -- & Different Depth & Single Task &  0.100 &   7.123 & 1.09E-12***     &  0.208 \\
Complex Environment & Simple Environment & -- & Different Depth & Dual Task   &  0.160 &  11.328 & 1.14E-29***     &  0.332 \\
\bottomrule
\end{tabular*}
\begin{center}
\footnotesize Significance: * \textit{p} < .05, ** \textit{p} < .01, *** \textit{p} < .001.
\end{center}
\end{table*}

\begin{table*}[hbp]
\centering
\footnotesize
\caption{Pairwise comparisons of estimated marginal means of log reaction times across repeated spatial configurations, separately for task type, virtual depth, and physical environmental complexity. Positive estimates indicate longer reaction times in Level~1 relative to Level~2.}
\label{tab:RT_Emmeans_Repeated}
\begin{tabular*}{\textwidth}{@{\extracolsep{\fill}}ccccccccc@{}}
\toprule
\multicolumn{2}{c}{\textbf{Comparison}} & \multicolumn{3}{c}{\textbf{Condition}} & \multicolumn{4}{c}{\textbf{Result}} \\
\cmidrule(lr){1-2} \cmidrule(lr){3-5} \cmidrule(lr){6-9}
\textbf{Level 1} & \textbf{Level 2} & \textbf{Environment} & \textbf{Depth} & \textbf{Task Type} & \textbf{Estimate} & \textbf{$t(23349)$} & \textbf{$P_{\text{adjusted}}$} & \textbf{Cohen's $d$} \\
\midrule
\multicolumn{9}{l}{\textit{Task type comparisons}} \\
Dual Task   & Single Task & Simple Environment  & Same Depth      & --          &  0.042 &  1.490 & 1.37E-01        &  0.087 \\
Dual Task   & Single Task & Simple Environment  & Different Depth & --          &  0.112 &  4.010 & 6.08E-05***     &  0.234 \\
Dual Task   & Single Task & Complex Environment & Same Depth      & --          &  0.047 &  1.640 & 1.00E-01        &  0.097 \\
Dual Task   & Single Task & Complex Environment & Different Depth & --          &  0.194 &  6.890 & 5.84E-12***     &  0.404 \\
\addlinespace
\multicolumn{9}{l}{\textit{Depth comparisons}} \\
Different Depth  & Same Depth & Simple Environment  & -- & Single Task &  0.110 &  3.904 & 9.49E-05***     &  0.228 \\
Different Depth  & Same Depth & Simple Environment  & -- & Dual Task   &  0.180 &  6.447 & 1.16E-10***     &  0.375 \\
Different Depth  & Same Depth & Complex Environment & -- & Single Task & -0.085 & -2.981 & 2.88E-03**      & -0.176 \\
Different Depth  & Same Depth & Complex Environment & -- & Dual Task   &  0.063 &  2.249 & 2.45E-02*       &  0.132 \\
\addlinespace
\multicolumn{9}{l}{\textit{Environment comparisons}} \\
Complex Environment & Simple Environment & -- & Same Depth      & Single Task &  0.219 &  7.746 & 9.90E-15***     &  0.456 \\
Complex Environment & Simple Environment & -- & Same Depth      & Dual Task   &  0.224 &  7.993 & 1.38E-15***     &  0.466 \\
Complex Environment & Simple Environment & -- & Different Depth & Single Task &  0.025 &  0.888 & 3.75E-01        &  0.052 \\
Complex Environment & Simple Environment & -- & Different Depth & Dual Task   &  0.107 &  3.810 & 1.40E-04***     &  0.222 \\
\bottomrule
\end{tabular*}
\begin{center}
\footnotesize Significance: * \textit{p} < .05, ** \textit{p} < .01, *** \textit{p} < .001.
\end{center}
\end{table*}

\begin{table*}[hbp]
\centering
\footnotesize
\caption{Pairwise comparisons of estimated marginal means for keypress accuracy across combined spatial configurations, separately for task type, virtual depth, and environmental complexity. Odds ratios greater than 1 indicate higher accuracy in Level~1 relative to Level~2.}
\label{tab:Accuracy Emmeans All}
\begin{tabular*}{\textwidth}{@{\extracolsep{\fill}}ccccccccc@{}}
\toprule
\multicolumn{2}{c}{\textbf{Comparison}} & \multicolumn{3}{c}{\textbf{Condition}} & \multicolumn{3}{c}{\textbf{Result}} \\
\cmidrule(lr){1-2} \cmidrule(lr){3-5} \cmidrule(lr){6-8}
\textbf{Level 1} & \textbf{Level 2} & \textbf{Environment} & \textbf{Depth} & \textbf{Task Type} & \textbf{Odds Ratio} & \textbf{$z$} & \textbf{$P_{\text{adjusted}}$} \\
\midrule
\multicolumn{8}{l}{\textit{Task type comparisons}} \\
Dual Task & Single Task & Simple Environment  & Same Depth      & --          & 1.295 &  1.100 & 2.71E-01 \\
Dual Task & Single Task & Complex Environment & Same Depth      & --          & 1.190 &  1.076 & 2.82E-01 \\
Dual Task & Single Task & Simple Environment  & Different Depth & --          & 0.816 & -0.885 & 3.76E-01 \\
Dual Task & Single Task & Complex Environment & Different Depth & --          & 0.930 & -0.409 & 6.83E-01 \\
\addlinespace
\multicolumn{8}{l}{\textit{Depth comparisons}} \\
Different Depth & Same Depth & Simple Environment  & -- & Single Task & 1.370 &  1.433 & 1.52E-01 \\
Different Depth & Same Depth & Complex Environment & -- & Single Task & 1.238 &  1.306 & 1.92E-01 \\
Different Depth & Same Depth & Simple Environment  & -- & Dual Task   & 0.863 & -0.602 & 5.47E-01 \\
Different Depth & Same Depth & Complex Environment & -- & Dual Task   & 0.968 & -0.186 & 8.52E-01 \\
\addlinespace
\multicolumn{8}{l}{\textit{Environment comparisons}} \\
Complex Environment & Simple Environment & -- & Same Depth      & Single Task & 0.540 & -3.491 & 4.81E-04*** \\
Complex Environment & Simple Environment & -- & Different Depth & Single Task & 0.488 & -3.428 & 6.07E-04*** \\
Complex Environment & Simple Environment & -- & Same Depth      & Dual Task   & 0.496 & -3.129 & 1.76E-03** \\
Complex Environment & Simple Environment & -- & Different Depth & Dual Task   & 0.556 & -2.912 & 3.60E-03** \\
\bottomrule
\end{tabular*}
\begin{center}
\footnotesize Significance: * \textit{p} < .05, ** \textit{p} < .01, *** \textit{p} < .001.
\end{center}
\end{table*}

\begin{table*}[hbp]
\centering
\footnotesize
\caption{Pairwise comparisons of estimated marginal means for keypress accuracy across novel spatial configurations, separately for task type, virtual depth, and environmental complexity. Odds ratios greater than 1 indicate higher accuracy in Level~1 relative to Level~2.}
\label{tab:Accuracy_Emmeans_Novel}
\begin{tabular*}{\textwidth}{@{\extracolsep{\fill}}ccccccccc@{}}
\toprule
\multicolumn{2}{c}{\textbf{Comparison}} & \multicolumn{3}{c}{\textbf{Condition}} & \multicolumn{3}{c}{\textbf{Result}} \\
\cmidrule(lr){1-2} \cmidrule(lr){3-5} \cmidrule(lr){6-8}
\textbf{Level 1} & \textbf{Level 2} & \textbf{Environment} & \textbf{Depth} & \textbf{Task Type} & \textbf{Odds Ratio} & \textbf{$z$} & \textbf{$P_{\text{adjusted}}$} \\
\midrule
\multicolumn{8}{l}{\textit{Task type comparisons}} \\
Dual Task   & Single Task & Simple Environment  & Same Depth      & --          & 0.742 & -1.502 & 1.33E-01 \\
Dual Task   & Single Task & Simple Environment  & Different Depth & --          & 0.708 & -1.574 & 1.15E-01 \\
Dual Task   & Single Task & Complex Environment & Same Depth      & --          & 0.969 & -0.205 & 8.38E-01 \\
Dual Task   & Single Task & Complex Environment & Different Depth & --          & 0.829 & -1.264 & 2.06E-01 \\
\addlinespace
\multicolumn{8}{l}{\textit{Depth comparisons}} \\
Different Depth  & Same Depth & Simple Environment  & -- & Single Task & 1.245 &  0.979 & 3.28E-01 \\
Different Depth  & Same Depth & Simple Environment  & -- & Dual Task   & 1.189 &  0.896 & 3.70E-01 \\
Different Depth  & Same Depth & Complex Environment & -- & Single Task & 0.990 & -0.064 & 9.49E-01 \\
Different Depth  & Same Depth & Complex Environment & -- & Dual Task   & 0.847 & -1.127 & 2.60E-01 \\
\addlinespace
\multicolumn{8}{l}{\textit{Environment comparisons}} \\
Complex Environment & Simple Environment & -- & Same Depth      & Single Task & 0.517 & -3.559 & 3.73E-04*** \\
Complex Environment & Simple Environment & -- & Different Depth & Single Task & 0.411 & -4.450 & 8.59E-06*** \\
Complex Environment & Simple Environment & -- & Same Depth      & Dual Task   & 0.675 & -2.318 & 2.05E-02* \\
Complex Environment & Simple Environment & -- & Different Depth & Dual Task   & 0.481 & -4.214 & 2.51E-05*** \\
\bottomrule
\end{tabular*}
\begin{center}
\footnotesize Significance: * \textit{p} < .05, ** \textit{p} < .01, *** \textit{p} < .001.
\end{center}
\end{table*}

\begin{table*}[hbp]
\centering
\footnotesize
\caption{Pairwise comparisons of estimated marginal means for keypress accuracy across repeated spatial configurations, separately for task type, virtual depth, and environmental complexity. Odds ratios greater than 1 indicate higher accuracy in Level~1 relative to Level~2.}
\label{tab:Accuracy_Emmeans_Repeated}
\begin{tabular*}{\textwidth}{@{\extracolsep{\fill}}ccccccccc@{}}
\toprule
\multicolumn{2}{c}{\textbf{Comparison}} & \multicolumn{3}{c}{\textbf{Condition}} & \multicolumn{3}{c}{\textbf{Result}} \\
\cmidrule(lr){1-2} \cmidrule(lr){3-5} \cmidrule(lr){6-8}
\textbf{Level 1} & \textbf{Level 2} & \textbf{Environment} & \textbf{Depth} & \textbf{Task Type} & \textbf{Odds Ratio} & \textbf{$z$} & \textbf{$P_{\text{adjusted}}$} \\
\midrule
\multicolumn{8}{l}{\textit{Task type comparisons}} \\
Dual Task   & Single Task & Simple Environment  & Same Depth      & --          & 2.262 &  1.931 & 5.35E-02 \\
Dual Task   & Single Task & Simple Environment  & Different Depth & --          & 0.940 & -0.155 & 8.77E-01 \\
Dual Task   & Single Task & Complex Environment & Same Depth      & --          & 1.462 &  1.337 & 1.81E-01 \\
Dual Task   & Single Task & Complex Environment & Different Depth & --          & 1.044 &  0.133 & 8.94E-01 \\
\addlinespace
\multicolumn{8}{l}{\textit{Depth comparisons}} \\
Different Depth  & Same Depth & Simple Environment  & -- & Single Task & 1.506 &  1.097 & 2.73E-01 \\
Different Depth  & Same Depth & Simple Environment  & -- & Dual Task   & 0.626 & -1.044 & 2.97E-01 \\
Different Depth  & Same Depth & Complex Environment & -- & Single Task & 1.548 &  1.520 & 1.28E-01 \\
Different Depth  & Same Depth & Complex Environment & -- & Dual Task   & 1.106 &  0.317 & 7.52E-01 \\
\addlinespace
\multicolumn{8}{l}{\textit{Environment comparisons}} \\
Complex Environment & Simple Environment & -- & Same Depth      & Single Task & 0.564 & -1.920 & 5.48E-02 \\
Complex Environment & Simple Environment & -- & Different Depth & Single Task & 0.579 & -1.491 & 1.36E-01 \\
Complex Environment & Simple Environment & -- & Same Depth      & Dual Task   & 0.364 & -2.442 & 1.46E-02* \\
Complex Environment & Simple Environment & -- & Different Depth & Dual Task   & 0.644 & -1.215 & 2.24E-01 \\
\bottomrule
\end{tabular*}
\begin{center}
\footnotesize Significance: * \textit{p} < .05, ** \textit{p} < .01, *** \textit{p} < .001.
\end{center}
\end{table*}

\clearpage

\section{Supplementary Tables of Perceived Workload}
\label{appendix:perceived_workload}

\begin{table}[hbp]
  \centering
  \setlength{\tabcolsep}{20pt}
  \caption{Random intercept variance and standard deviation at the participant level for cumulative link mixed models fitted to each NASA--TLX subscale.}
  \label{tab:clmm_random_effects}
  \begin{tabular}{lcc}
    \toprule
    \textbf{Subscale} & \textbf{Variance} & \textbf{Standard Deviation} \\
    \midrule
    Mental Demand    & 4.508   &   2.213 \\
    Physical Demand  & 23.866  &   4.885 \\
    Temporal Demand  & 5.102   &   2.258 \\
    Performance      & 6.748   &   2.597 \\
    Effort           & 2.978   &   1.726 \\
    Frustration      & 20.159  &   4.490 \\
    \bottomrule
  \end{tabular}
\end{table}

\begin{table*}[hbp]
\centering
\footnotesize
\caption{Mean and Standard deviation (within parenthesis) of the NASA-TLX scores for each sub-scale grouped by Condition. The dual-task condition significantly increased subjective workload across all condition}
\label{tab:nasa_tlx_by_condition_compact}
\begin{tabular*}{\textwidth}{@{\extracolsep{\fill}}ccccccccc@{}}
\toprule
\multicolumn{3}{c}{\textbf{Condition}} & \multicolumn{6}{c}{\textbf{NASA--TLX (Mean (SD))}} \\
\cmidrule(lr){1-3}\cmidrule(lr){4-9}
\textbf{Environment} & \textbf{Depth} & \textbf{Task Type} & \textbf{Mental Demand} & \textbf{Physical Demand} & \textbf{Temporal Demand} & \textbf{Performance$^{\dagger}$} & \textbf{Effort} & \textbf{Frustration} \\
\midrule
Simple Environment  & Same Depth      & Single Task & 39.5 (24.7) & 28.5 (22.6) & 28.0 (20.9) & 31.3 (22.1) & 34.2 (20.9) & 18.0 (17.1) \\
Simple Environment  & Same Depth      & Dual Task   & 62.8 (21.3) & 40.8 (26.7) & 51.2 (26.3) & 42.8 (20.7) & 62.3 (19.9) & 34.3 (24.3) \\
Simple Environment  & Different Depth & Single Task & 42.0 (21.8) & 31.9 (23.7) & 27.8 (21.0) & 35.0 (20.1) & 39.1 (18.8) & 24.7 (19.1) \\
Simple Environment  & Different Depth & Dual Task   & 60.1 (24.0) & 40.8 (26.8) & 52.7 (27.4) & 43.5 (21.8) & 58.0 (22.3) & 36.2 (26.7) \\
Complex Environment & Same Depth      & Single Task & 46.5 (24.4) & 39.2 (26.7) & 33.4 (21.8) & 38.0 (22.1) & 42.5 (22.3) & 31.5 (25.1) \\
Complex Environment & Same Depth      & Dual Task   & 69.3 (19.3) & 47.3 (31.6) & 50.5 (25.4) & 48.1 (21.4) & 68.8 (18.0) & 43.1 (29.7) \\
Complex Environment & Different Depth & Single Task & 54.8 (21.6) & 42.0 (26.4) & 33.9 (21.8) & 40.1 (17.1) & 48.6 (21.9) & 33.4 (24.9) \\
Complex Environment & Different Depth & Dual Task   & 74.4 (16.9) & 49.1 (31.2) & 53.3 (26.4) & 48.5 (21.8) & 71.8 (15.7) & 41.2 (28.2) \\
\bottomrule
\end{tabular*}
\begin{minipage}{\textwidth}
 \footnotesize \textbf{Note.} The $^{\dagger}$ \textit{Performance} sub-scale is labelled from `Perfect' to `Failure', i.e., a lower score is better.
\end{minipage}
\end{table*}

\clearpage

\section{Supplementary Tables of Qualitative Results}
\label{appendix:interview_results}
To complement the quantitative results reported in the main text, the following tables present selected excerpts from the interview data.

\begin{table*}[hbp]
\centering
\caption{Participants reported the easiest condition for finding the target ``T''. The simple environment with same depth under single-task conditions received the highest number of responses.}
\label{tab:Interview_FindT}
\setlength{\tabcolsep}{20pt}
\begin{tabular}{cccc}
\toprule
\multicolumn{3}{c}{\textbf{Condition}} & \multicolumn{1}{c}{\textbf{Participants}} \\
\cmidrule(lr){1-3}\cmidrule(lr){4-4}
\textbf{Environment} & \textbf{Depth} & \textbf{Task Type} & \textbf{N} \\
\midrule
Simple Environment  & Same Depth      & Single Task & 21 \\
Simple Environment  & Same Depth      & Dual Task   & 0 \\
Simple Environment  & Different Depth & Single Task & 3 \\
Simple Environment  & Different Depth & Dual Task   & 0 \\
Complex Environment & Same Depth      & Single Task & 0 \\
Complex Environment & Same Depth      & Dual Task   & 0 \\
Complex Environment & Different Depth & Single Task & 0 \\
Complex Environment & Different Depth & Dual Task   & 0 \\
\bottomrule
\end{tabular}
\begin{center}
\footnotesize
Total responses: 24 participants.
\end{center}
\end{table*}

\begin{table*}[hbp]
\centering
\caption{Participants reported the hardest condition for finding the target T. The complex environment with different depths under dual-task conditions received the highest number of responses.}
\label{tab:Interview_Hardest_FindT}
\setlength{\tabcolsep}{20pt}
\begin{tabular}{cccc}
\toprule
\multicolumn{3}{c}{\textbf{Condition}} & \multicolumn{1}{c}{\textbf{Participants}} \\
\cmidrule(lr){1-3}\cmidrule(lr){4-4}
\textbf{Environment} & \textbf{Depth} & \textbf{Task Type} & \textbf{N} \\
\midrule
Simple Environment  & Same Depth      & Single Task & 0 \\
Simple Environment  & Same Depth      & Dual Task   & 0 \\
Simple Environment  & Different Depth & Single Task & 0 \\
Simple Environment  & Different Depth & Dual Task   & 0 \\
Complex Environment & Same Depth      & Single Task & 0 \\
Complex Environment & Same Depth      & Dual Task   & 6 \\
Complex Environment & Different Depth & Single Task & 1 \\
Complex Environment & Different Depth & Dual Task   & 17 \\
\bottomrule
\end{tabular}
\begin{center}
\footnotesize
Total responses: 24 participants.
\end{center}
\end{table*}

\begin{table*}[hbp]
\centering
\caption{Participants reported the independent variables with the greatest impact on their ability to find the target T. Environment was selected most frequently.}
\label{tab:Interview_IV_FindT}
\setlength{\tabcolsep}{20pt}
\begin{tabular}{cc}
\toprule
\textbf{Independent Variable} & \textbf{Participants (N)} \\
\midrule
Environment      & 14 \\
Depth            & 1 \\
Task             & 9 \\
\bottomrule
\end{tabular}
\begin{center}
\footnotesize
Total responses: 24 participants.
\end{center}
\end{table*}

\begin{table*}[hbp]
\centering
\caption{Participants reported the easiest condition for remembering repeated spatial configurations. The simple environment with same depth under single-task conditions received the highest number of responses.}
\label{tab:Interview_Easiest}
\setlength{\tabcolsep}{20pt}
\begin{tabular}{cccc}
\toprule
\multicolumn{3}{c}{\textbf{Condition}} & \multicolumn{1}{c}{\textbf{Participants}} \\
\cmidrule(lr){1-3}\cmidrule(lr){4-4}
\textbf{Environment} & \textbf{Depth} & \textbf{Task Type} & \textbf{N} \\
\midrule
Simple Environment  & Same Depth      & Single Task & 13 \\
Simple Environment  & Same Depth      & Dual Task   & 0 \\
Simple Environment  & Different Depth & Single Task & 7 \\
Simple Environment  & Different Depth & Dual Task   & 0 \\
Complex Environment & Same Depth      & Single Task & 0 \\
Complex Environment & Same Depth      & Dual Task   & 0 \\
Complex Environment & Different Depth & Single Task & 4 \\
Complex Environment & Different Depth & Dual Task   & 0 \\
\bottomrule
\end{tabular}
\begin{center}
\footnotesize
Total responses: 24 participants.
\end{center}
\end{table*}

\begin{table*}[hbp]
\centering
\caption{Participants reported the hardest condition for remembering repeated spatial configurations. The complex environment with different depths under dual-task conditions received the highest number of responses.}
\label{tab:Interview_Hardest}
\setlength{\tabcolsep}{20pt}
\begin{tabular}{cccc}
\toprule
\multicolumn{3}{c}{\textbf{Condition}} & \multicolumn{1}{c}{\textbf{Participants}} \\
\cmidrule(lr){1-3}\cmidrule(lr){4-4}
\textbf{Environment} & \textbf{Depth} & \textbf{Task Type} & \textbf{N} \\
\midrule
Simple Environment  & Same Depth      & Single Task & 0 \\
Simple Environment  & Same Depth      & Dual Task   & 4 \\
Simple Environment  & Different Depth & Single Task & 0 \\
Simple Environment  & Different Depth & Dual Task   & 0 \\
Complex Environment & Same Depth      & Single Task & 0 \\
Complex Environment & Same Depth      & Dual Task   & 6 \\
Complex Environment & Different Depth & Single Task & 1 \\
Complex Environment & Different Depth & Dual Task   & 13 \\
\bottomrule
\end{tabular}
\begin{center}
\footnotesize
Total responses: 24 participants.
\end{center}
\end{table*}

\begin{table*}[hbp]
\centering
\caption{Participants reported the independent variables with the greatest impact on memory for repeated spatial configurations. Task type was selected most frequently.}
\label{tab:Interview_IV_Impact}
\setlength{\tabcolsep}{20pt}
\begin{tabular}{cc}
\toprule
\textbf{Independent Variable} & \textbf{Participants (N)} \\
\midrule
Environment      & 8 \\
Depth            & 1 \\
Task Type        & 15 \\
\bottomrule
\end{tabular}
\begin{center}
\footnotesize
Total responses: 24 participants.
\end{center}
\end{table*}

\clearpage

\section{Additional Figures}
\label{appendix:one-tailed}

\begin{figure*}[hbp]
    \centering
    \includegraphics[width=0.98\textwidth]{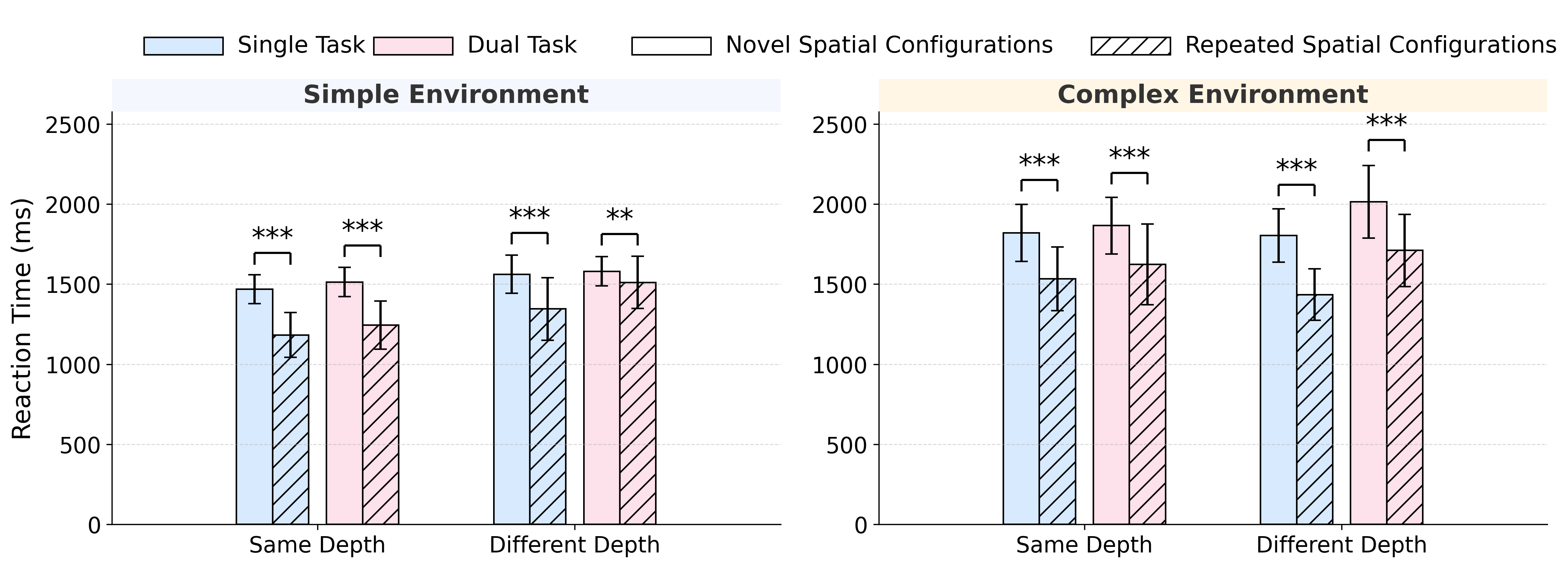}
    \caption{Bar plots of the mean reaction times across task type, environment, and depth conditions for novel and repeated spatial configurations. Statistical significance was assessed using pairwise comparisons of estimated marginal means for novel versus repeated spatial configurations, with Tukey-adjusted p-values. Asterisks denote significant pairwise comparisons (* \textit{p} < .05, ** \textit{p} < .01, *** \textit{p} < .001).}
    \Description{This figure shows bar plots of the mean reaction times to our contextual cueing task. Reaction times are shown separately for novel and repeated spatial configurations across task type (single vs. dual task), environment (simple vs. complex), and depth (same vs. different depth). Reaction time was measured in milliseconds from the onset of the target to the participant's correct response, with incorrect trials excluded from the averages. Statistical significance was assessed using pairwise comparisons of estimated marginal means for novel versus repeated spatial configurations, with Tukey-adjusted p-values. Asterisks denote significant pairwise comparisons (* \textit{p} < .05, ** \textit{p} < .01, *** \textit{p} < .001).}
    \label{fig:One-Tailed}
\end{figure*}

\end{document}